\documentclass{article}

\usepackage{tikz, pgf, pgfplots}
\usetikzlibrary
	{
		arrows, automata, chains, datavisualization.formats.functions, fit, positioning, quantikz, quotes, shapes
	}

\usepackage{amsthm}
\usepackage{amssymb}
\usepackage[bb = boondox]{mathalfa}
\usepackage{dsfont}
\usepackage{mathtools}
\usepackage{multicol}
\usepackage{braket}
\usepackage{physics}
\numberwithin{equation}{section}

\usepackage{yquant, braket}
\usetikzlibrary{quantikz, fit}

\usepackage[a4paper, left = 2.5cm, right = 2.5 cm, top = 2.5cm, bottom = 3.0 cm]{geometry}

\usepackage{xcolor}
\usepackage{varwidth}
\usepackage[breakable, skins, theorems]{tcolorbox}
\usepackage{graphicx}
\usepackage{hyperref}
\hypersetup
{
	colorlinks,
	linkcolor = {WordRed!75!black},
	citecolor = {WordBlueDarker25},
	urlcolor = {WordAquaDarker50}
}
\usepackage{nicematrix}
\NiceMatrixOptions
{
	code-for-first-row = \color{RedPurple},
	code-for-last-col = \color{WordAquaDarker25}
}

\usepackage{colortbl}
\usepackage{float}
\usepackage{cellspace}
\usepackage{makecell}
\usepackage[boxed, ruled, vlined]{algorithm2e}
\usepackage{appendix}
\usepackage{multirow}
\newcommand{\orcidicon}[1]{\href{https://orcid.org/#1}{\includegraphics[height=\fontcharht\font`\B]{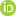}}}

\newtheorem{definition}{Definition}[section]

\definecolor{MyLightRed}{RGB}{244, 213, 245}
\definecolor{WordRed}{RGB}{255, 0, 102}
\definecolor{RedDarkLightest}{HTML}{ff0088}
\definecolor{RedDarkLight}{HTML}{ea005f}
\definecolor{RedPurple}{HTML}{aa007f}
\definecolor{Purple}{HTML}{911146}
\definecolor{WordLightGreen}{RGB}{140, 214, 192}
\definecolor{WordGreen}{RGB}{0, 176, 80}
\definecolor{GreenLightest}{HTML}{00ffa0}
\definecolor{GreenLighter1}{HTML}{00b383}
\definecolor{GreenLighter2}{HTML}{00aa7f}
\definecolor{GreenDark}{HTML}{225522}
\definecolor{GreenTeal}{HTML}{008080}
\definecolor{WordIceBlue}{RGB}{223, 227, 229}
\definecolor{MyVeryLightBlue}{RGB}{211, 245, 247}
\definecolor{WordBlueVeryLight}{RGB}{0, 176, 240}
\definecolor{WordBlueLight}{RGB}{0, 112, 192}
\definecolor{WordBlueDark}{RGB}{46, 116, 181}
\definecolor{WordBlueDarker}{RGB}{31, 78, 121}
\definecolor{WordBlueDarker25}{RGB}{54, 96, 146}
\definecolor{WordBlueDarker50}{RGB}{36, 64, 98}
\definecolor{WordBlueDarkest}{RGB}{0, 32, 96}
\definecolor{WordBlue}{RGB}{19, 65, 99}
\definecolor{MyBlue}{RGB}{0, 64, 128}
\definecolor{MyDarkBlue}{RGB}{0, 51, 102}
\definecolor{BlueVeryDark}{HTML}{222255}
\definecolor{WordAquaLighter80}{RGB}{218, 238, 243}
\definecolor{WordAquaLighter60}{RGB}{183, 222, 232}
\definecolor{WordAquaLighter40}{RGB}{146, 205, 220}
\definecolor{WordAquaDarker25}{RGB}{49, 134, 155}
\definecolor{WordAquaDarker50}{RGB}{33, 89, 103}
\definecolor{WordVeryLightTeal}{RGB}{223, 236, 235}
\definecolor{WordLightTeal}{RGB}{160, 199, 197}
\definecolor{WordDarkTealLighter80}{RGB}{207, 223, 234}
\definecolor{WordDarkTeal}{RGB}{72, 123, 119}
\definecolor{WordDarkerTeal}{RGB}{48, 82, 80}
\definecolor{WordTurquoiseLighter80}{RGB}{209, 238, 249}

\title{A two-party quantum parliament}

\author{
	Theodore Andronikos$^1$\orcidicon{0000-0002-3741-1271}
	and
	Michael Stefanidakis$^1$\orcidicon{0000-0001-7077-1079}\\
	$^1$Department of Informatics, Ionian University, \\
	7 Tsirigoti Square, 49100 Corfu, Greece; \\
	\{andronikos, mistral\}@ionio.gr \\
}

\begin{document}

\maketitle

\begin{abstract}
	This paper introduces the first functional model of a quantum parliament that is dominated by two parties or coalitions, and may or may not contain independent legislators. We identify a single crucial parameter, aptly named \emph{free will radius}, that can be used as a practical measure of the quantumness of the parties and the parliament as a whole. The free will radius used by the two parties determines the degree of independence that is afforded to the representatives of the parties. Setting the free will radius to zero degrades the quantum parliament to a classical one. On the other hand, setting the free will radius to its maximum value $1$, makes the representatives totally independent. Moreover, we present a quantum circuit with which we simulate in Qiskit the operation of the quantum parliament under various scenarios. The experimental results allow to arrive at some novel and fundamental conclusions that, we believe, provide new insights on the operation and the traits of a possible future quantum parliament. Finally, we propose the novel game of ``Passing the Bill,'' which captures the operation of the quantum parliament and basic options available to the leadership of the two parties.

	\textbf{Keywords:} Quantum parliament; quantum legislators; quantum simulation; qiskit; quantum voting; quantum games.
\end{abstract}

\section{Introduction} \label{sec:Introduction}

In the current era, two crucial ingredients of the typical political system are the parliament and the political parties or political coalitions. The parliament is formed ny a fixed number $N$ of legislators or representatives that most often are members of political parties and, occasionally, are independent, in the sense that they have no ties to any specific party. Periodically, elections are held to renew the parliament. It is generally acknowledged that this type of a political system, despite its strengths and advantages, has ample room for improvement. This work discusses the prospects of improving, even slightly, the current party and parliament system, by switching to some form of quantum parliament. Furthermore, we give some preliminary evidence regarding the usefulness of independent representatives, in terms of their potential to improve the efficiency of the parliamentary institution.

One might reasonably ask why consider a quantum and not just a probabilistic parliament. Apart from the fact that probabilistic political systems have already been studied, there ia a deeper reason. The quantum era is only a few years away. Without attempting to predict the precise time length to the beginning of the quantum era, the prevalent feeling is that it will bring fundamental changes  to our computational abilities and overall way of life. The field of quantum information and quantum computation took some time to achieve its present status; its modest origins can be placed in the early '80s. Today there are already available commercial quantum computers. In the near future the available quantum computers will be much more powerful, employing significantly more qubits. The increase in the number of qubits will enhance our computational capabilities to the point where it will be feasible to solve practical instances of difficult decision and optimization problems, which are just inaccessible today with our classical technology. Quantum cryptography, quantum internet, and quantum machine learning will radically affect many aspects of the future society. Under this emerging new paradigm, it seems natural to ask whether concepts and tools from the quantum realm can provide any advantage to our political system.

\subsection{Related work}

One of the first proposal for the of a quantum system of government centered around a quantum parliament can be found in \cite{Aerts2005}. In this original work, it is argued compellingly that some form of quantum democracy can enhance the present widely established democratic political system and perhaps even remedy some of its shortcomings.
This can hopefully be achieved by introducing applying ideas and notions from quantum mechanics to the operation of the present-day political system.

Obviously quantum voting is a crucial part of any quantum parliament. It not surprising then that most previous works have examined important issues regarding the implementation of the voting process. To be viable, any quantum voting protocol must satisfy certain critical properties. Many researchers believe that these properties are: (1) privacy, (2) non-reusability, (3) verifiability, (4) fairness, (5) eligibility, and (6) correctness.

A quantum protocol that can transmit classical bits and qubits anonymously broadcast was proposed as early as in 2005 by Christandl and Wehner \cite{Christandl2005}. Hillery et al. in 2006  \cite{Hillery2006} introduced the anonymous traveling ballot. The following year, a quantum anonymous voting protocol of the traveling ballot scheme was proposed in \cite{Vaccaro2007}. Later, \cite{Li2008} demonstrated a quantum anonymous voting system based on the entangled state, where one can vote for multiple candidates. \cite{Wang2013} contains a quantum voting protocol that uses a non-symmetric quantum channel and teleportation. Efficient use of teleportation that enhances the security of the voting process, is also involved in the quantum voting protocol presented by the authors in \cite{Tian2015}. A voting scheme based on quantum proxy blind signature was given in \cite{Zhang2017a}. A simple, yet efficient, quantum voting scheme involving pairs of particles in a multi-particle entangled state, and which can be used for large scale voting, was demonstrated in \cite{Xue2017}. An interesting voting protocol based on quantum Blockchain, that additionally satisfies the self-tallying property, was shown in \cite{Sun2018}. A fault-tolerant quantum anonymous voting protocol was presented in \cite{Wang2019}. \cite{Zhang2019} introduced a novel quantum anonymous voting protocol that succeeds in protecting the both the voting content of the voters and the number of the votes received by the candidate. Very recently, \cite{Gao2021}, proposed a new quantum election protocol that, besides fulfilling the typical requirements, it also satisfies the receipt-freeness property. An interesting critique regarding quantum voting protocols is found in \cite{Arapinis2021}, highlighting the lack of rigor in terms of definitions and proofs used to ascertain their security and efficiency.


In view of above references, one may surmise that the critical technical issues involved in quantum voting protocols have been addressed in a more or less satisfactory way by many researchers. For this reason we shall assume that when it comes to the security and privacy of the act of voting some of the proposed solutions have been implemented and have addressed these legitimate concerns. These issues are undoubtedly important. It is our belief, however, that even if/when they are satisfactorily solved, the main question of the political process and how it may be improved to benefit ordinary people will persist. In this paper our focus is entirely on the possibility of high level manipulation of a quantum parliament, especially in the typical case of two party political systems. Our rationale behind our approach is the following: politics is most often about outcomes that affect the social gain and not really about technical scientific issues that lie in the field of quantum cryptography. When the technical issues have been solved, the fundamental issue of policies that can be harmful or beneficial for the public will remain. So our main line of inquiry is the following: if in the future the political systems switch to forms of quantum democracy and establish quantum parliaments, will this bring some positive changes to the people, or, unfortunately, nothing essential will change? Politics is not in its essence about technical mathematical issues, it's about actions and reforms that may affect negatively or positively our way of life. Can the coming quantum era bring some hope for improvement? In that perspective, this work is similar in spirit with that of \cite{Pluchino2011}.


In this paper we propose the first, to the best of our knowledge, functional model of a quantum parliament. The parliament is comprised primarily of two parties, as well as a number of independent legislators. In our theoretical analysis of the proposed model we identify a single crucial parameter, aptly named \emph{free will radius}, that can be used as a practical measure of the quantumness of the parties and the parliament as a whole. The free will radius used by the two parties determines the degree of independence that is afforded to the representatives of the parties. Setting the free will radius to zero degrades the quantum parliament to a classical one. On the other hand, setting the free will radius to its maximum value $1$, makes the representatives totally independent. Moreover, we present a quantum circuit with which we simulate in Qiskit the operation of the quantum parliament under various scenarios. The experimental results allow to arrive at some novel and fundamental conclusions that, we believe, provide new insights on the operation and the traits of a possible future quantum parliament. We briefly mention the most important of our conclusions below.

The degree of independence of the representatives is strictly increasing with the free will radius. As the free will radius increases, the legislators become more independent, to the point where they are completely independent when the free will radius becomes $1$. Then the legislators have total freedom to decide how to vote. We may go as far as to say that free will maters more than numbers, in the sense that the majority party may still lose a vote, if its representatives are given a high enough free will radius. This is even more evident when the two parties employ different free will radii. The odds improve drastically for the party that employs the smaller free will radius. The party that grants the greater freedom to its representatives is more prone to lose the vote. The existence or not of independent representatives in a quantum parliament is of secondary importance. For small values of the free will radii the independent legislators just favor the majority party. For higher values of the free will radii their presence is less discernible. Their presence becomes important only when the two parties have the same number of representatives. In such a case they may provide a significant impetus to the voting procedure, by turning an impossibility to a probable eventuality. Such a situation can only happen when the parties employ zero or very small free will radii. When the free will radii increase, the effect of the independent legislators becomes negligible. Therefore, in an almost classical parliament where two parties have the same number of representatives their presence is useful and may even deemed necessary. However, in a true quantum parliament their value is diminished. These observations have motivated us to propose the novel game of ``Passing the Bill,'' which captures the operation of the quantum parliament and the dilemmata posed to the leadership of the two parties.

\subsection{Organization}

This paper is structured as follows. Section \ref{sec:Introduction} gives an introduction to the subject along with the most relevant references. Section \ref{sec:The Setting} introduces and explains the assumptions used for the formulation of our model. Section \ref{sec:Analyzing the Voting Process} contains the mathematical analysis of the model. Section \ref{sec:Simulating the Quantum Parliament} presents the experimental data obtained from our simulations. Section \ref{sec:The Game of Passing the Bill} describes the game of ``Passing the Bill,'' and section \ref{sec:Discussion & Conclusions} provides a brief summary and discussion of the results.

\section{The setting} \label{sec:The Setting}

\subsection{The classical setting} \label{subsec:The Classical Setting}

Perhaps the most prevalent political system today is the so called ``representative democracy.'' In such a system, the voters periodically (typically every four or five years) choose in free and secret elections a fixed number $N$ of representatives, also refereed to as legislators. These $N$ representatives usually belong to political parties or political coalitions, that stand for specific beliefs and values. It is also quite conceivable that some representatives are independent, having no association or ties with any particular party. The $N$ representatives   comprise the parliament, the political entity holding the legislative power, which is responsible for passing bills that are supposed, at least in theory, to advance the social welfare of the people. Bills are debated in the parliament and are accepted as laws when
the majority of the representatives vote in favor. There also exist rare cases, where a greater percentage of votes is necessary, when the bill under question induces constitutional amendments. This is desirable because of the special role of the constitution in modern democracies.

\subsection{The quantum setting} \label{subsec:The Quantum Setting}

In this subsection we describe the quantum parliament we envision and state clearly the assumptions underlying its operation. Let us point out before hand that many important issues will not be addressed in this work. Undoubtedly, the most critical issue left out is the possibility of quantum general elections. We shall take for granted that elections, either of a classical type, or of an unspecified quantum type have resulted in the formation of a given quantum parliament. For ease of exposition we assume that the quantum parliament consists of two parties and we further consider the case where, apart from the two parties, there are also independent representatives.

Below we list the assumptions, the terminology and the notation used throughout this paper.

\begin{itemize}
	\item[\textbf{A1:}]	There are two political parties/coalitions, Alice's party and Bob's party, designated by $A$ and $B$, respectively.
	\item[\textbf{A2:}]	$n_A$ denotes the number of representatives from Alice's party, $n_B$ the number of representatives from Bob's party, and $n_I$ the number of independent representatives. Obviously
						\begin{align} \label{eq:Distribution of Representatives with Independent}
							n_A + n_B + n_I = N \ .
						\end{align}
	If there are no independent representatives, the above equation  (\ref{eq:Distribution of Representatives with Independent}) simplifies to
						\begin{align} \label{eq:Distribution of Representatives without Independent}
							n_A + n_B = N \ .
						\end{align}
	\item[\textbf{A3:}]	For every bill, both parties decide whether to vote in favor or against, i.e., \emph{yes} or \emph{no}. The decision made by both parties is ``classical'' in the sense that is an absolute yes or no. This being a quantum parliament means that every legislator casts a \emph{quantum} vote. Hence, each vote is represented by a \emph{qubit} in the $2$-dimensional Hilbert space $\mathcal{H}_{2}$ and the underlying orthonormal basis is taken to be $H = \{ \ket{ yes }, \ket{ no } \}$. The intuition behind this convention is straightforward: $\ket{ yes }$ and $\ket{ no }$ correspond to the classical decisions \emph{yes} and \emph{no}, respectively.
	\item[\textbf{A4:}]	As seems fitting in a quantum parliament, this party decision is viewed as a \emph{suggestion} or \emph{recommendation} from the leadership of the party to its representatives. Specifically, both parties $A$ and $B$ adopt and tolerate \emph{free will} radii $r_A$ and $r_B$, respectively, such that $0 \leq r_A, r_B < 1$. This is viewed as a result of the democratic tradition and sociopolitical culture of each party. Therefore, the vote $\ket{ \psi }$ of a legislator need not coincide with the decision of the party's leadership; it suffices to be within a \emph{distance} less than or equal to the free will radius of the party. The notion of \emph{distance} will be made rigorous in the next section \ref{sec:Simulating the Quantum Parliament}.
	\item[\textbf{A5:}]	The above assumptions describe the typical operation of the quantum parliament. However, we allow for exceptional circumstances, in which one or both parties view the bill in question to be of great importance, and to succeed in passing or stopping the bill, they decide to temporarily abolish their corresponding free will radius. This forces their representatives to obey the party decision, but, at the same time this behavior, being deeply autocratic and undemocratic, incurs a heavy political and social cost. Such situations will be examined in section \ref{sec:The Game of Passing the Bill}.
	\item[\textbf{A6:}]	In case there are independent legislators, these are unaffected by the decisions of the parties. As a result, they behave according to their conscience and their vote is abstracted by the standard qubit representation formula $a \ket{ yes } + b \ket{ no }, \ a, b \in \mathbb{C}, \ |a|^2 + |b|^2 = 1$.
\end{itemize}

Thus, in view of assumption \textbf{A4}, all representatives of the same party must adhere the same free will radius when voting, but the distance of each one's vote from the party decision will be different. In contrast to party representatives, independent legislators enjoy total freedom to vote as they deem fit. Let us clarify that the above model does not involve entanglement in any way. The decision making process of each legislator is completely independent of the respective decision of her fellow legislators, both party affiliated and independent. The party representatives simply follow the official party line within the imposed distance constraint. Furthermore, the measurement of the whole voting process is performed at the end, after all legislators have cast their vote. This seems more appropriate, since measuring each vote after being cast would severely degrade the quantumness of the system.

\section{Analyzing the voting process} \label{sec:Analyzing the Voting Process}

In every parliamentarian system there are many occasions where both parties and the independent legislators agree to vote in favor of a bill that, in their view, advances social welfare. In such cases, the voting process becomes a rather typical formality, as the result is predetermined in advance. Such cases will be of no concern in this paper. Instead, we shall focus on cases where the two parties, and the independent representatives, if they exist, are in strong disagreement regarding a specific bill. In such situations the voting process can be viewed as a game between the two parties. We pursue this perspective in section \ref{sec:The Game of Passing the Bill}.

\begin{figure}[H]
	\begin{tcolorbox}
		[
			colback = gray!03,
			enhanced jigsaw, 
			sharp corners,
			boxrule = 0.25 pt,
			sharp corners,
			center title,
			fonttitle = \bfseries
		]
		\centering
		\begin{tikzpicture}[scale = 1.0]
			\def \angle { 360 / 12 }
			\def \radius { 3 cm }
			\draw [ thick, ->, > = { Latex [round] } ] ( - \radius - 2 cm, 0 ) -- ( \radius + 2 cm, 0 ) node [ right ] { $y$ };
			\draw [ thick, ->, > = { Latex [round] } ] ( 0,  - \radius - 2 cm ) -- ( 0,  \radius + 2 cm ) node [ above ] { $z$ };
			\draw [ very thick, ->, > = { Latex [round] }, RedPurple ] ( 0.0, 0.0 ) -- ( { \radius * cos(3 * \angle) }, { \radius * sin(3 * \angle) } ) node [ above = 3.5 mm, right ] { $\ket{yes}$ } node [ above = 3.5 mm, left ] { A };
			\draw [ very thick, ->, > = { Latex [round] }, GreenLighter2 ] ( 0.0, 0.0 ) -- ( { \radius * cos(9 * \angle) }, { \radius * sin(9 * \angle) } ) node [ below = 3.5 mm, right ] { $\ket{no}$ } node [ below = 3.5 mm, left ] { B };
			\draw [ thin, RedPurple ] ( 0.0, 0.0 ) -- ( { \radius * cos(2.0 * \angle) }, { \radius * sin(2.0 * \angle) } );
			\draw [ thin, RedPurple ] ( 0.0, 0.0 ) -- ( { \radius * cos(4.0 * \angle) }, { \radius * sin(4.0 * \angle) } );
			\draw [ thin, GreenLighter2 ] ( 0.0, 0.0 ) -- ( { \radius * cos(8.0 * \angle) }, { \radius * sin(8.0 * \angle) } );
			\draw [ thin, GreenLighter2 ] ( 0.0, 0.0 ) -- ( { \radius * cos(10.0 * \angle) }, { \radius * sin(10.0 * \angle) } );
			\begin{scope}[on background layer]
				\draw [thick, WordBlueDark] ( 0, 0 ) circle ( \radius );
				\draw [ thick, RedPurple, dotted ] ( 0, \radius ) circle ( 0.5 * \radius );
				\draw [ <->, > = { Latex [round] }, RedPurple, line width = 1.0 mm ] ( { \radius * cos(2.0 * \angle) }, { \radius * sin(2.0 * \angle) } ) arc ( 2 * \angle : 4 * \angle : 3 );
				\draw [ ->, > = { Latex [round] }, RedPurple, line width = 0.3 mm ] ( { 1.7 * \radius * cos(4.0 * \angle) }, { 1.7 * \radius * sin(4.0 * \angle) } ) node [ RedPurple, left ] { $r_A$ } -- ( { 1.04 * \radius * cos(3.5 * \angle ) }, { 1.04 * \radius * sin(3.5 * \angle) } );
				\draw [ thick, GreenLighter2, dotted ] ( 0, - \radius ) circle ( 0.5 * \radius );
				\draw [ <->, > = { Latex [round] }, GreenLighter2, line width = 1.0 mm ] ( { \radius * cos(8.0 * \angle) }, { \radius * sin(8.0 * \angle) } ) arc ( 8 * \angle : 10 * \angle : 3 );
				\draw [ ->, GreenLighter2, line width = 0.3 mm ] ( { 1.7 * \radius * cos(8.0 * \angle) }, { 1.7 * \radius * sin(8.0 * \angle) } ) node [ GreenLighter2, left ] { $r_B$ } -- ( { 1.04 * \radius * cos(8.5 * \angle ) }, { 1.04 * \radius * sin(8.5 * \angle) } );
			\end{scope}
		\end{tikzpicture}
		\caption{This figure provides a $2$-dimensional visual representation of the free will radii. Representatives from party A need not vote $\ket{yes}$, but their vote must be within a distance of $r_A$ from $\ket{yes}$, that is on the shaded red portion of the circle. Symmetrically, legislators from party B may not vote $\ket{no}$, but their vote must be within a distance of $r_B$ from $\ket{no}$, that is on the shaded green portion of the circle.}
		\label{fig:Free Will Radii of Party A and Party B}
	\end{tcolorbox}
\end{figure}
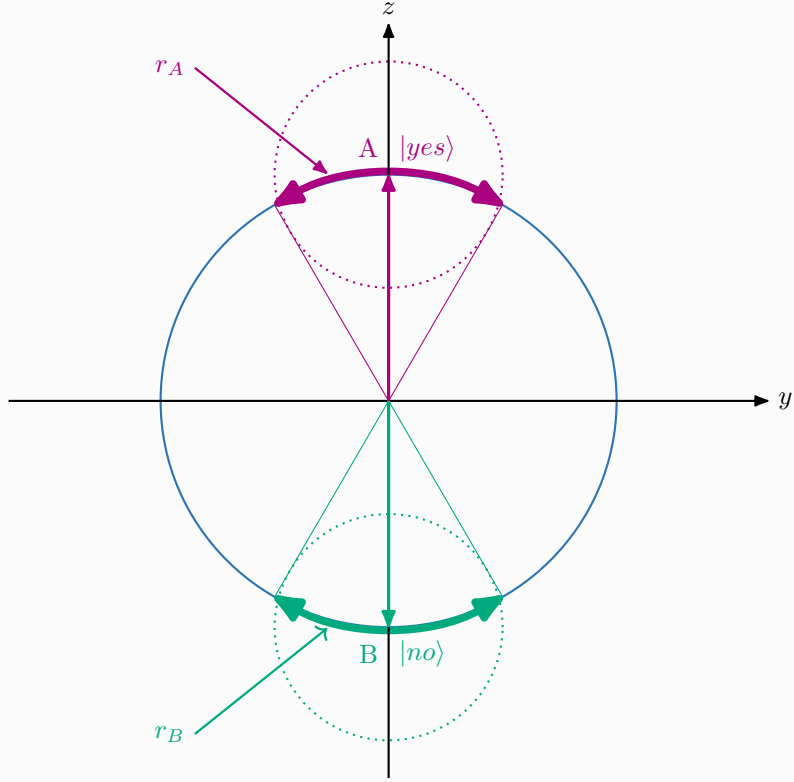

Without loss of generality, we shall assume that Alice's party $A$ is in favor of the bill, whereas Bob's party $B$ is against the bill. The opposite situation can be treated in an entirely symmetric way. Figure \ref{fig:Free Will Radii of Party A and Party B} provides a simplified $2$-dimensional visual representation of the situation concerning the free will radii. It can be considered as a projection of the Bloch sphere onto the $yz$ plane. Representatives from party A need not vote $\ket{yes}$, but their vote must be within a distance of $r_A$ from $\ket{yes}$, that is on the shaded red portion of the circle. In terms of the Bloch sphere, their vote must lie on the Bloch sphere and at the same time be withing a virtual sphere centered at $\ket{yes}$ and having radius $r_A$. Symmetrically, legislators from party B may not vote $\ket{no}$, but their vote must be within a distance of $r_B$ from $\ket{no}$, that is on the shaded green portion of the circle. In the complete $3$-dimensional picture, their vote is a point on the Bloch sphere, which also happens to be inside a virtual sphere centered at $\ket{no}$ and having radius $r_B$.

The representatives of Alice's party are allowed some room for maneuver, which practically means that their vote, viewed as a quantum state, need not be identical with the party recommendation, which, according to our previous assumptions, is $\ket{ 0 }$. Therefore, if $\ket{ \psi_j^A }$ is the quantum state corresponding to the vote of representative $j, 1 \leq j \leq n_A$, the distance between the states $\ket{ 0 }$ and $\ket{ \psi_j^A }$ must be less than party's A free will radius $r_A$. In a symmetrical fashion, if $\ket{ \psi_k^B }$ is the quantum state corresponding to the vote of representative $k, 1 \leq k \leq n_B$, of Bob's party, the distance between the states $\ket{ 1 }$ and $\ket{ \psi_k^B }$ must be less than $r_B$. A standard metric used to determine how close are two quantum states are is the \emph{trace distance} (see for more details the books by \cite{Nielsen2010} and \cite{Wilde2018}).

\begin{definition}[Trace distance] \label{def:Trace Distance}
	The \emph{trace distance} between two quantum states represented by their density matrices $\rho$ and $\sigma$ respectively, is computed by the next formula
	\begin{align} \label{eq:Trace Distance}
		d_{\rm tr} ( \rho, \sigma ) = \frac{ 1 }{ 2 } \trace \abs{ \rho - \sigma } \ .
	\end{align}
\end{definition}

A quick reminder is perhaps in order here: given an operator $U$, $U^\dagger$ is its \emph{adjoint}, its \emph{trace} $\trace U$ is the sum of its diagonal elements, that is $\trace U = \sum_{ i } \expval{ U }{ i }$, and $\abs{ U }$ is defined as $\abs{ U } = \sqrt{ U^\dagger U}$, i.e., as the positive square root of $U^\dagger U$.

The general form of the density matrix $\rho_{ \psi }$ corresponding to a pure state $\ket { \psi } = a \ket{ 0 } + b \ket{ 1 }$, where $\abs{ a }^{ 2 } + \abs{ b }^{ 2 } = 1$, is given by the next formula

\begin{align} \label{eq:Density Matrix of a Pure State}
	\rho_{ \psi } = \dyad{ \psi }
	=
	\begin{bNiceMatrix}[ margin ] 
		\abs{ a }^{ 2 } & a b^{ \star }
		\\
		b a^{ \star } & \abs{ b }^{ 2 }
	\end{bNiceMatrix}
	\ .
\end{align}

In this case, $\rho_{ \psi }$ is the \emph{orthogonal projector} onto the unit ket $\ket { \psi }$. In the special cases where $\ket { \psi } = \ket { 0 }$ and $\ket { \psi } = \ket { 1 }$, the corresponding density matrices are

\begin{center}
	\begin{minipage}[b]{0.45 \textwidth}
		\begin{align} \label{eq:Density Matrix of Ket 0}
			\rho_{ 0 } = \dyad{ 0 }
			=
			\begin{bNiceMatrix}[ margin ] 
				1 & 0
				\\
				0 & 0
			\end{bNiceMatrix}
		\end{align}
	\end{minipage}
	\begin{minipage}[b]{0.45 \textwidth}
		\begin{align} \label{eq:Density Matrix of Ket 1}
			\rho_{ 1 } = \dyad{ 1 }
			=
			\begin{bNiceMatrix}[ margin ] 
				0 & 0
				\\
				0 & 1
			\end{bNiceMatrix}
			\ .
		\end{align}
	\end{minipage}
\end{center}

By combining (\ref{eq:Density Matrix of Ket 0}) and (\ref{eq:Density Matrix of Ket 1}) with (\ref{eq:Density Matrix of a Pure State}), we derive

\begin{align} \label{eq:Trace Distance from Ket 0 - I}
	\rho_{ 0 } - \rho_{ \psi }
	=
	\begin{bNiceMatrix}[ margin ] 
		1 - \abs{ a }^{ 2 } & - a b^{ \star }
		\\
		- b a^{ \star } & - \abs{ b }^{ 2 }
	\end{bNiceMatrix}
	=
	\left( \rho_{ 0 } - \rho_{ \psi } \right)^{\dagger}
\end{align}

and

\begin{align} \label{eq:Trace Distance from Ket 1 - I}
	\rho_{ 1 } - \rho_{ \psi }
	=
	\begin{bNiceMatrix}[ margin ] 
		- \abs{ a }^{ 2 } & - a b^{ \star }
		\\
		- b a^{ \star } & 1 - \abs{ b }^{ 2 }
	\end{bNiceMatrix}
	=
	\left( \rho_{ 0 } - \rho_{ \psi } \right)^{\dagger}
	\ ,
\end{align}

respectively. Taking advantage of the fact that $\abs{ a }^{ 2 } + \abs{ b }^{ 2 } = 1$, we may simplify (\ref{eq:Trace Distance from Ket 0 - I}) and (\ref{eq:Trace Distance from Ket 1 - I}) as follows

\begin{center}
	\begin{minipage}[b]{0.4 \textwidth}
		\begin{align} \label{eq:Trace Distance from Ket 0 - II}
			\rho_{ 0 } - \rho_{ \psi }
			=
			\begin{bNiceMatrix}[ margin ] 
				\abs{ b }^{ 2 } & - a b^{ \star }
				\\
				- b a^{ \star } & - \abs{ b }^{ 2 }
			\end{bNiceMatrix}
		\end{align}
	\end{minipage}
	\begin{minipage}[b]{0.4 \textwidth}
		\begin{align} \label{eq:Trace Distance from Ket 1 - II}
			\rho_{ 1 } - \rho_{ \psi }
			=
			\begin{bNiceMatrix}[ margin ] 
				- \abs{ a }^{ 2 } & - a b^{ \star }
				\\
				- b a^{ \star } & \abs{ a }^{ 2 }
			\end{bNiceMatrix}
			\ .
		\end{align}
	\end{minipage}
\end{center}

It is now easy to see that

\begin{align} \label{eq:Trace Distance from Ket 0 - III}
	\left( \rho_{ 0 } - \rho_{ \psi } \right)^{\dagger}
	\left( \rho_{ 0 } - \rho_{ \psi } \right)
	=
	\begin{bNiceMatrix}
		\abs{ b }^{ 4 } + \abs{ a }^{ 2 } \abs{ b }^{ 2 } & - a b^{ \star } \abs{ b }^{ 2 } + a b^{ \star } \abs{ b }^{ 2 }
		\\
		- b a^{ \star } \abs{ b }^{ 2 } + b a^{ \star } \abs{ b }^{ 2 } & \abs{ a }^{ 2 } \abs{ b }^{ 2 } + \abs{ b }^{ 4 }
	\end{bNiceMatrix}
	=
	\begin{bNiceMatrix}[ margin ] 
		\abs{ b }^{ 2 } & 0
		\\
		0 & \abs{ b }^{ 2 }
	\end{bNiceMatrix}
\end{align}

and

\begin{align} \label{eq:Trace Distance from Ket 1 - III}
	\left( \rho_{ 1 } - \rho_{ \psi } \right)^{\dagger}
	\left( \rho_{ 1 } - \rho_{ \psi } \right)
	=
	\begin{bNiceMatrix}
		\abs{ a }^{ 4 } + \abs{ a }^{ 2 } \abs{ b }^{ 2 } & a b^{ \star } \abs{ a }^{ 2 } - a b^{ \star } \abs{ a }^{ 2 }
		\\
		b a^{ \star } \abs{ a }^{ 2 } - b a^{ \star } \abs{ a }^{ 2 } & \abs{ a }^{ 2 } \abs{ b }^{ 2 } + \abs{ a }^{ 4 }
	\end{bNiceMatrix}
	=
	\begin{bNiceMatrix}[ margin ] 
		\abs{ a }^{ 2 } & 0
		\\
		0 & \abs{ a }^{ 2 }
	\end{bNiceMatrix}
	\ .
\end{align}

There is a general result (consult \cite{Nielsen2010} p.$75$) that if $U = \sum_{ i } a_{ i } \dyad{ i }$ is the spectral decomposition of the Hermitian operator $U$, then

\begin{align} \label{eq:Operator Function}
	f ( U )
	=
	\sum_{ i } f ( a_{ i } ) \dyad{ i }
	\ .
\end{align}

If $U$ is also \emph{positive}, meaning its eigenvalues $a_{ i }$ are nonnegative, then, in the special case where $f ( U ) = \sqrt{ U }$, (\ref{eq:Operator Function}) becomes

\begin{align} \label{eq:Square Root Operator Function}
	\sqrt{ U }
	=
	\sum_{ i } \sqrt{ a_{ i } } \dyad{ i }
	\ .
\end{align}

Although density matrices are Hermitian and positive, the important observation in Definition \ref{def:Trace Distance} is that for any operator $U$, $U^\dagger U$ is positive (see for instance \cite{Nielsen2010}, p.$71$). Hence, we may apply (\ref{eq:Square Root Operator Function}) and derive that

\begin{align} \label{eq:Norm Ket 0 - Ket psi}
	\abs{ \rho_{ 0 } - \rho_{ \psi } }
	=
	\left(
	\left( \rho_{ 0 } - \rho_{ \psi } \right)^{\dagger}
	\left( \rho_{ 0 } - \rho_{ \psi } \right)
	\right)^{ \frac{1}{2} }
	\overset{ ( \ref{eq:Trace Distance from Ket 0 - III} ) }{ = }
	\begin{bNiceMatrix}[ margin ] 
		\abs{ b } & 0
		\\
		0 & \abs{ b }
	\end{bNiceMatrix}
	\ ,
\end{align}

and

\begin{align} \label{eq:Norm Ket 1 - Ket psi}
	\abs{ \rho_{ 1 } - \rho_{ \psi } }
	=
	\left(
	\left( \rho_{ 1 } - \rho_{ \psi } \right)^{\dagger}
	\left( \rho_{ 1 } - \rho_{ \psi } \right)
	\right)^{ \frac{1}{2} }
	\overset{ ( \ref{eq:Trace Distance from Ket 1 - III} ) }{ = }
	\begin{bNiceMatrix}[ margin ] 
		\abs{ a } & 0
		\\
		0 & \abs{ a }
	\end{bNiceMatrix}
	\ .
\end{align}

Finally, we compute the trace distance between the basis kets $\ket{ 0 }$, $\ket{ 1 }$ and the pure state $\ket { \psi }$

\begin{align} \label{eq:Trace Distance from Ket 0}
	d_{\rm tr} ( \rho_{ 0 }, \rho_{ \psi } )
	=
	\frac{ 1 }{ 2 } \trace \abs{ \rho_{ 0 } - \rho_{ \psi } }
	\overset{ ( \ref{eq:Norm Ket 0 - Ket psi} ) }{ = }
	\abs{ b }
	\ ,
\end{align}

and

\begin{align} \label{eq:Trace Distance from Ket 1}
	d_{\rm tr} ( \rho_{ 1 }, \rho_{ \psi } )
	=
	\frac{ 1 }{ 2 } \trace \abs{ \rho_{ 1 } - \rho_{ \psi } }
	\overset{ ( \ref{eq:Norm Ket 1 - Ket psi} ) }{ = }
	\abs{ a }
	\ .
\end{align}

A typical way to represent any pure state $\ket { \psi }$ as a point on the Bloch sphere  is to write

\begin{align} \label{eq:Bloch Sphere Representation of a Pure State}
	\ket{ \psi }
	=
	\cos \frac{ \theta }{ 2 }
	\ket{ 0 }
	+
	e^{ i \varphi } \sin \frac{ \theta }{ 2 }
	\ket{ 1 }
	\ , \
	0 \leq \theta \leq \pi
	,
	0 \leq \varphi \leq 2 \pi
	\ .
\end{align}

In the above representation (\ref{eq:Bloch Sphere Representation of a Pure State}), angles $\theta$ and $\varphi$ are the elevation and azimuth angles respectively of the point $\ket { \psi }$ of the Bloch sphere. In view of (\ref{eq:Bloch Sphere Representation of a Pure State}), (\ref{eq:Trace Distance from Ket 0}) and (\ref{eq:Trace Distance from Ket 1}) become

\begin{center}
	\begin{minipage}[b]{0.45 \textwidth}
		\begin{align} \label{eq:Bloch Sphere Representation of Trace Distance from Ket 0}
			d_{\rm tr} ( \rho_{ 0 }, \rho_{ \psi } )
			=
			\sin \frac{ \theta }{ 2 }
		\end{align}
	\end{minipage}
	\begin{minipage}[b]{0.45 \textwidth}
		\begin{align} \label{eq:Bloch Sphere Representation of Trace Distance from Ket 1}
			d_{\rm tr} ( \rho_{ 1 }, \rho_{ \psi } )
			=
			\cos \frac{ \theta }{ 2 }
		\end{align}
	\end{minipage}
\end{center}

If $\rho_{ \psi_j^A }$ is the density matrix corresponding to the vote $\ket{ \psi_j^A } = \cos \frac{ \theta_j^A }{ 2 } \ket{ 0 } + e^{ i \varphi_j^A } \sin \frac{ \theta_j^A }{ 2 } \ket{ 1 }$ of representative $j, 1 \leq j \leq n_A$, of Alice's party, then, according to (\ref{eq:Bloch Sphere Representation of Trace Distance from Ket 0}), the requirement that $\ket{ \psi_j^A }$ lies inside the \emph{free will} sphere, with center $\ket{ 0 }$ and radius $r_A$, is captured by the next constraint

\begin{align} \label{eq:Requirement for Vote Inside Free Will Sphere from Ket 0}
	d_{\rm tr} ( \rho_{ 0 }, \rho_{ \psi_j^A } )
	=
	\sin \frac{ \theta_j^A }{ 2 } \leq r_A
	\Rightarrow
	\theta_j^A \leq 2 \arcsin r_A
	\ , \
	0 \leq \theta_j^A \leq \pi
	\ .
\end{align}

Symmetrically, if $\rho_{ \psi_k^B }$ is the density matrix corresponding to the vote $\ket{ \psi_k^B } = \cos \frac{ \theta_k^B }{ 2 } \ket{ 0 } + e^{ i \varphi_k^B } \sin \frac{ \theta_k^B }{ 2 } \ket{ 1 }$ of representative $k, 1 \leq j \leq n_B$, of Bob's party, then, according to (\ref{eq:Bloch Sphere Representation of Trace Distance from Ket 1}), the requirement that $\ket{ \psi_k^B }$ lies inside the \emph{free will} sphere, with center $\ket{ 1 }$ and radius $r_B$, is equivalent to

\begin{align} \label{eq:Requirement for Vote Inside Free Will Sphere from Ket 1}
	d_{\rm tr} ( \rho_{ 1 }, \rho_{ \psi_k^B } )
	=
	\cos \frac{ \theta_k^B }{ 2 } \leq r_B
	\Rightarrow
	\theta_k^B \leq 2 \arccos r_B
	\ , \
	0 \leq \theta_k^B \leq \pi
	\ .
\end{align}

Thus, every legislator from Alice's party has to adhere to the party rules and vote so that requirement (\ref{eq:Requirement for Vote Inside Free Will Sphere from Ket 0}) is satisfied. Analogously, every representative from Bob's party must vote in such a way that (\ref{eq:Requirement for Vote Inside Free Will Sphere from Ket 1}) is fulfilled. Independent legislator, when they exist, have complete freedom to do as they deem fit, and their vote is abstracted by the most general formula (\ref{eq:Bloch Sphere Representation of a Pure State}). This situation is summarized in Table \ref{tbl:Voting Process Constraints}.

\begin{table}[H]
	\centering
	\renewcommand{\arraystretch}{2.0}
	\begin{tcolorbox}
		[
			colback = gray!03,
			enhanced jigsaw, 
			sharp corners,
			boxrule = 0.1 pt,
			toprule = 0.1 pt,
			bottomrule = 0.1 pt
		]
		\caption{This Table contains the constraints governing the vote of every representative, according to her party affiliation. \\}
		\label{tbl:Voting Process Constraints}
		\begin{tabular}{ >{\centering\arraybackslash} m{7.5 cm} !{\vrule width 1.25 pt} >{\centering\arraybackslash} m{6.5 cm} }
			\Xhline{4\arrayrulewidth}
			\multicolumn{2}{c}{The voting process}
			\\
			\Xhline{\arrayrulewidth}
			Party affiliation of legislator
			&
			Constraint
			\\
			\Xhline{3\arrayrulewidth}
			Alice's party: $\ket{ \psi_j^A } = \cos \frac{ \theta_j^A }{ 2 } \ket{ 0 } + e^{ i \varphi_j^A } \sin \frac{ \theta_j^A }{ 2 } \ket{ 1 }$
			&
			$\theta_j^A \leq 2 \arcsin r_A \ , \ 0 \leq \theta_j^A \leq \pi, 0 \leq \varphi \leq 2 \pi$
			\\
			\Xhline{\arrayrulewidth}
			Bob's party: $\ket{ \psi_k^B } = \cos \frac{ \theta_k^B }{ 2 } \ket{ 0 } + e^{ i \varphi_k^B } \sin \frac{ \theta_k^B }{ 2 } \ket{ 1 }$
			&
			$\theta_k^B \leq 2 \arccos r_B \ , \ 0 \leq \theta_k^B \leq \pi, 0 \leq \varphi \leq 2 \pi$
			\\
			\Xhline{\arrayrulewidth}
			Independent: $\ket{ \psi } = \cos \frac{ \theta }{ 2 } \ket{ 0 } + e^{ i \varphi } \sin \frac{ \theta }{ 2 } \ket{ 1 }$
			&
			Unconstrained: $0 \leq \theta \leq \pi, 0 \leq \varphi \leq 2 \pi$
			\\
			\Xhline{4\arrayrulewidth}
		\end{tabular}
	\end{tcolorbox}
	\renewcommand{\arraystretch}{1.0}
\end{table}

\section{Simulating the quantum parliament} \label{sec:Simulating the Quantum Parliament}

This virtual quantum parliament can be simulated with the help of Qiskit \cite{Qiskit2022}. Each vote is effectively a single qubit. It is reasonable to assume that legislators from Alice's party start with their preliminary vote in state $\ket{ yes }$, according to the decision of the leadership of their party. They arrive at their final vote after considering a number of factor, such as the overall political environment, their own ideology, and the free will radius allowed by the party. The situation is pretty much the same for the representatives of Bob's party, with the exception that their starting position is the state $\ket{ no }$. The synthetic process that each legislator carries out silently in her mind in order to determine her final vote, can be faithfully abstracted by a single qubit operator. Each representative is a different individual, and as such applies a different unitary operator to her initial state that leads to her ultimate decision and vote. Simulating this process, inevitably involves simulating arbitrary single qubit operator. This is rather easy to achieve relatively easy in modern quantum computers. One approach might be to invoke the fact that every unitary single qubit operator can be written in terms of simpler rotation operators followed by a phase shift (see standard references such as \cite{Nielsen2010}, \cite{Williams2010}, or \cite{Kaye2007} for a detailed analysis). These rotation and phase shift operators are realized as quantum gates in all contemporary quantum computers that employ the quantum gate model. Specifically, given any unitary single qubit operator $U$, there $\alpha, \beta, \gamma, \delta \in \mathbb{R}$ such that

\begin{align} \label{eq:Z-Y Decomposition of Single Qubit Operators}
	U = e^{ i \alpha } R_{ z } ( \beta ) R_{ y } ( \gamma ) R_{ z } ( \delta ) \ ,
\end{align}

where

\begin{center}
	\begin{minipage}[b]{0.45 \textwidth}
		\begin{align} \label{eq:Ry Rotation Operator}
			R_{ y } ( \theta )
			=
			\begin{bNiceMatrix}[ margin ] 
				\cos \frac{ \theta }{ 2 } & - \sin \frac{ \theta }{ 2 }
				\\
				\sin \frac{ \theta }{ 2 } & \phantom{-} \cos \frac{ \theta }{ 2 }
			\end{bNiceMatrix}
		\end{align}
	\end{minipage}
	\begin{minipage}[b]{0.45 \textwidth}
		\begin{align} \label{eq:Rz Rotation Operator}
			R_{ z } ( \theta )
			=
			\begin{bNiceMatrix}[ margin ] 
				e^{ - \frac{ i \theta }{ 2 } } & 0
				\\
				0 & e^{ \frac{ i \theta }{ 2 } }
			\end{bNiceMatrix}
			\ .
		\end{align}
	\end{minipage}
\end{center}

So we could simulate the voting process of the legislators in the quantum parliament by using the quantum gates $R_{ y } ( \theta )$ and $R_{ z } ( \theta )$ with the appropriate parameters. These operators rotate the state of the qubit around the $y$ and $z$ axes of the Bloch sphere. Another equivalent possibility, and the one we have opted for in our implementation is to directly use the general single qubit gate $U$, since all other rotation gates can be derived from $U$. Below we write the matrix corresponding to $U$ as given in the Qiskit documentation \cite{QiskitUGate2022}:

\begin{align} \label{eq:Matrix Representation of the U Gate}
	U ( \theta, \varphi, \lambda )
	=
	\begin{bmatrix}
		\cos \frac{\theta}{2} & - e^{ i \lambda } \sin \frac{\theta}{2}
		\\
		e^{ i \varphi } \sin \frac{\theta}{2} & e^{ i ( \varphi + \lambda ) } \cos \frac{\theta}{2}
	\end{bmatrix}
	\ .
\end{align}

In (\ref{eq:Matrix Representation of the U Gate}), which is in agreement with (\ref{eq:Bloch Sphere Representation of a Pure State}), parameters $\theta$ and $\varphi$ refer to the elevation and azimuth angles, respectively, of the Bloch sphere. Parameter $\lambda$ corresponds to a global phase factor that is physically unobservable. Of course, it goes without saying, that in our simulation the parameters of the $U$ gate are randomly generated and obey the constraints outlined in the previous section \ref{sec:Analyzing the Voting Process} and summarized in Table \ref{tbl:Voting Process Constraints}.

We simulate the operation of the quantum parliament in a typical situation, where, a political system is dominated by two parties (or coalitions), and one of them has the majority of the seats in the parliament. In our setting this is Alice's party. Let us denote by $p$ the probability that Alice's party succeeds in passing the bill. We experimentally estimate the probability $p$ under various scenarios, such as

\begin{itemize}
	\item	when Alice's and Bob's party both employ the same free will radii,
	\item	when the two parties have different free will radii, and
	\item	when independent legislators are also present in the parliament.
\end{itemize}

\begin{figure}[H]
	\begin{tcolorbox}
		[
			colback = gray!03,
			enhanced jigsaw, 
			sharp corners,
			boxrule = 0.25 pt,
			sharp corners,
			center title,
			fonttitle = \bfseries
		]
		\centering
		\includegraphics[scale = 0.35, trim = {0 0 0cm 0}, clip]{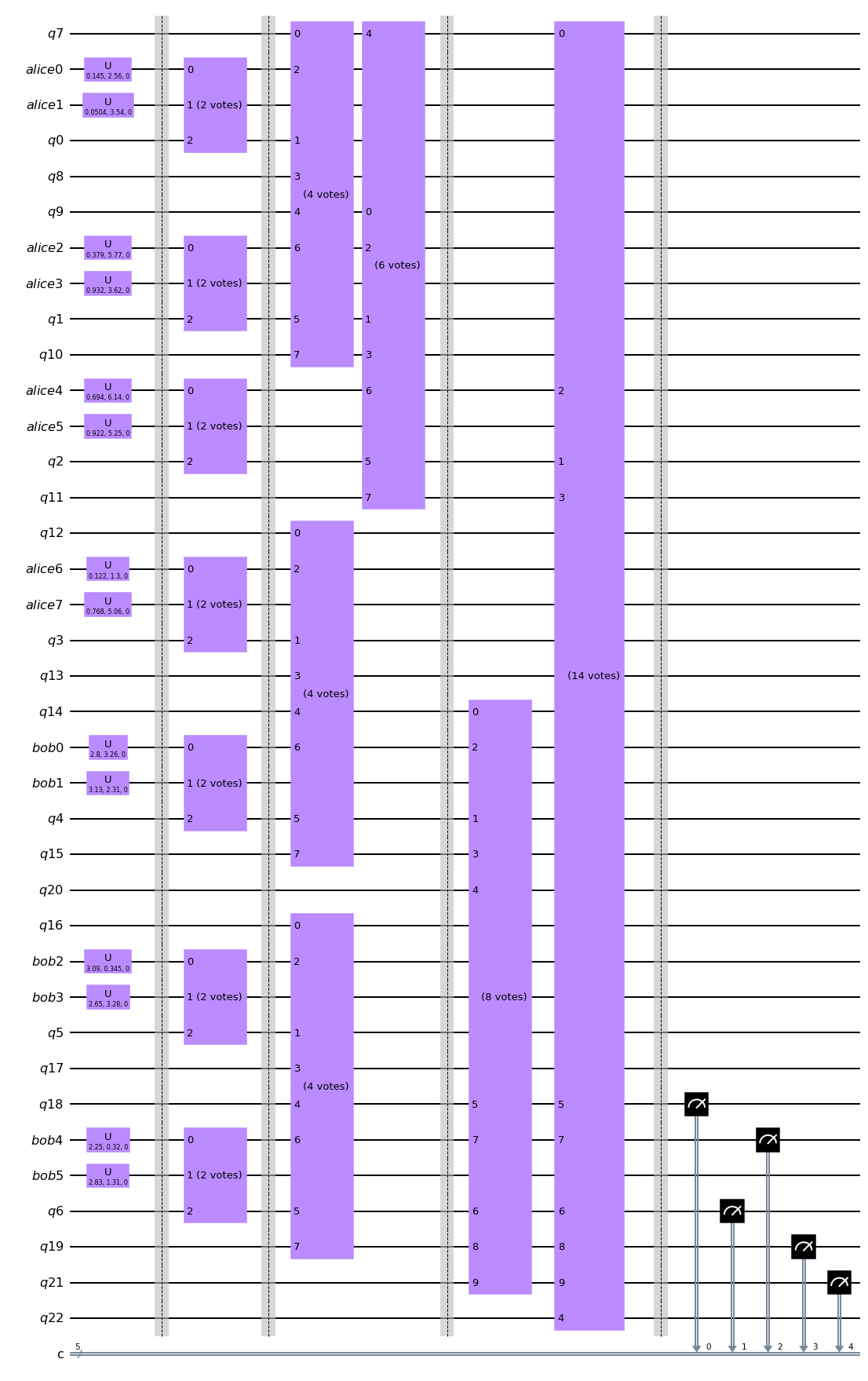}
		\caption{The circuit for simulating the quantum voting process in a parliament of 14 legislators, where Alice and Bob's parties have $8$ and $6$ representatives respectively. }
		\label{fig:A Quantum Parliament with 14 Representatives}
	\end{tcolorbox}
\end{figure}

The core of the Qiskit circuit for simulating the quantum voting process in a parliament of $14$ representatives is depicted in Figure \ref{fig:A Quantum Parliament with 14 Representatives}. In the circuit there are $14$ \emph{voter} qubits and $23$ \emph{ancilla} qubits, which are necessary in order to build a tree of adders that sum the final outcome. The latter is measured at the end of the experiment and mapped on $5$ classical qubits.

As mentioned previously, legislators from Alice's party are represented by qubits in state $\ket{yes}$ and correspond to positive votes, while legislators from Bob's party are in state $\ket{no}$ and correspond to negative votes.

The summation tree is built in a hierarchical manner. From left to right in Figure \ref{fig:A Quantum Parliament with 14 Representatives}:

\begin{itemize}
	\item	A \emph{voter initialization} layer applies the $U ( \theta, \varphi, \lambda )$ gate to each voter qubit, in order to initialize the state representing the decision of the voter.
	\item	A layer of \emph{$2$ voter adders} sums the votes in pairs of voters. Each $2$-vote subcircuit takes as input two qubits corresponding to votes plus an ancilla output qubit, and produces a $3$-qubit result in standard signed binary representation.
	\item	Subsequent layers combine the partial results from previous layers, presenting their outcome to the next layer's adder in an increasing number of qubits, capable of storing the intermediate and final vote counts. All binary adders in these layers are using the DraperQFTAdder circuit of Qiskit \cite{DraperQFTAdder2022}.
	\item	At the end, a measurement of the $5$ qubits representing the cumulative decision of all $14$ voters is taken.
\end{itemize}

Figure \ref{fig:A Quantum Parliament with 14 Representatives} provides a typical schematic for one of the many quantum circuits that were used to run the experiments. In this particular circuit one may see that Alice's party has 8 representatives, Bob's party has 6, there are no independent legislators, and both parties use the same free will radius, which, in this case, is $0.5$.

\subsection{The free will radius determines independence} \label{subsec:The Free Will Radius determines Independence}

In the first scenario Alice's party holds the majority having elected $8$ representatives, Bob's party is the opposition party with $6$ representatives, there are no independent legislators, and both parties have equal free will radii, i.e., $n_A = 8$, $n_B = 6$, $n_I = 0$, and $r_A = r_B$.

\begin{figure}[H]
	\begin{tcolorbox}
		[
			grow to left by = 1.0 cm,
			grow to right by = 1.0 cm,
			colback = gray!03,
			enhanced jigsaw, 
			sharp corners,
			toprule = 1.0 pt,
			bottomrule = 1.0 pt,
			leftrule = 0.1 pt,
			rightrule = 0.1 pt,
			sharp corners,
			center title,
			fonttitle = \bfseries
		]
		\centering
		\begin{minipage}{0.45 \textwidth}
			\includegraphics[scale = 0.45, trim = {0 0 0cm 0}, clip]{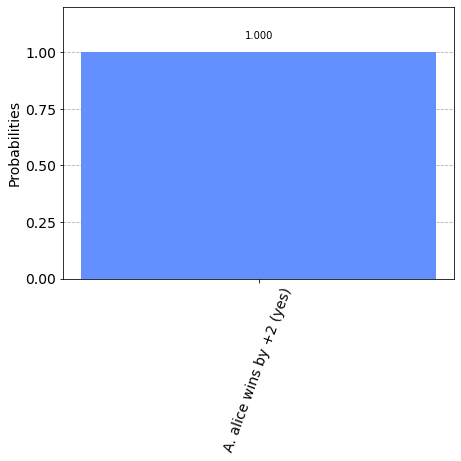}
			\caption{The case where both parties behave classically, by setting their free will radii to $0$. In this case the parliament is classical and the voting outcome is certain with probability $1.0$.}
			\label{fig:8+6+EqualFWR-0.0-MeasurementOutcomes}
		\end{minipage}
		\hfill
		\begin{minipage}{0.45 \textwidth}
			\includegraphics[scale = 0.45, trim = {0 0 0cm 0}, clip]{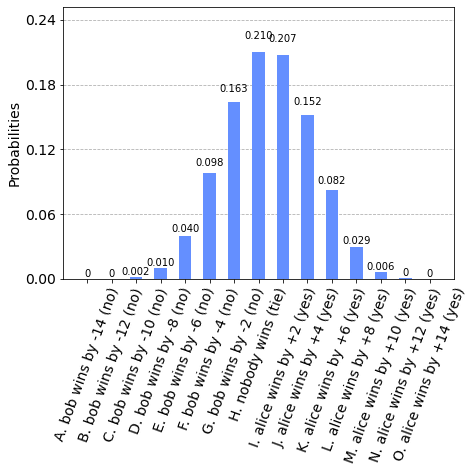}
			\caption{The possible outcomes and their probabilities for the same maximum free will radius $1$. In this case all representatives from both parties behave as independent legislators.}
			\label{fig:8+6+EqualFWR-1.0-MeasurementOutcomes}
		\end{minipage}
	\end{tcolorbox}
\end{figure}

We begin by considering the two extreme cases, which will provide us with further insight.

\begin{itemize}
	\item	If both parties set their free will radii to zero, that is $r_A = r_B = 0$, then the parliament degenerates to a classical one, where both parties function as is, more or less,  typical in most contemporary parliamentary democracies. Of course, this situation is of little interest, since every voting outcome is completely predictable and the majority always passes the bill with probability $p = 1.0$. This case, is given as a point of reference, in order to appreciate the operation of a true quantum parliament, as will be illustrated in the forthcoming figures. The final measurement, after the voting process is completed, will produce the unique predetermined outcome shown in Figure \ref{fig:8+6+EqualFWR-0.0-MeasurementOutcomes} with probability $1.0$.
	\item	The other extreme scenario is when both parties allow complete freedom of choice to their representatives, by setting their free will radii to the maximum value: $r_A = r_B = 1$. As demonstrated by the experimental results of Figure \ref{fig:8+6+EqualFWR-1.0-MeasurementOutcomes}, anything is possible now, albeit with different probabilities. The behavior of such a parliament is statistically indistinguishable from that of a parliament where all legislators are independent. As anticipated, a free will radius of $1$ means, in effect, that the party representatives behave as independent representatives. In view of the majority required to pass a bill, the odds of Alice passing the bill are now strictly less than $0.5$.
\end{itemize}

We clarify that the final outcome, shown in the figures above, combines both positive and negative votes and reports the \emph{difference} by which a specific party (Alice's or Bob's) wins.

\begin{figure}[H]
	\begin{tcolorbox}
		[
			grow to left by = 1.0 cm,
			grow to right by = 1.0 cm,
			colback = gray!03,
			enhanced jigsaw, 
			sharp corners,
			toprule = 1.0 pt,
			bottomrule = 1.0 pt,
			leftrule = 0.1 pt,
			rightrule = 0.1 pt,
			sharp corners,
			center title,
			fonttitle = \bfseries
		]
		\centering
		\begin{minipage}{0.45 \textwidth}
			\includegraphics[scale = 0.45, trim = {0 0 0cm 0}, clip]{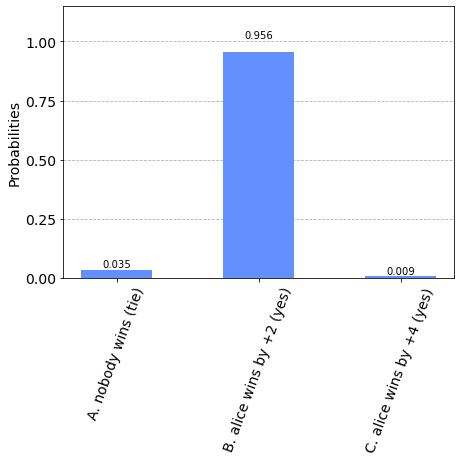}
			\caption{The possible outcomes and their probabilities when $r_A = r_B = 0.1$.}
			\label{fig:8+6+EqualFWR-0.1-MeasurementOutcomes}
		\end{minipage}
		\hfill
		\begin{minipage}{0.45 \textwidth}
			\includegraphics[scale = 0.45, trim = {0 0 0cm 0}, clip]{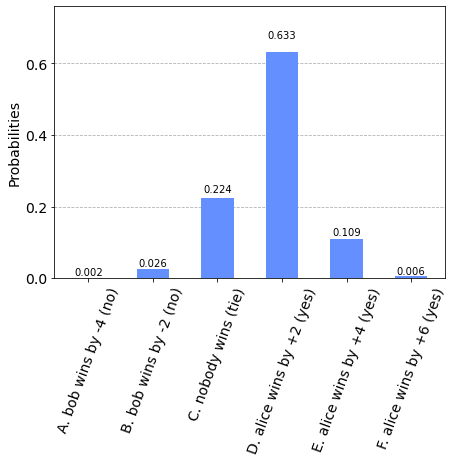}
			\caption{The possible outcomes and their probabilities when $r_A = r_B = 0.3$.}
			\label{fig:8+6+EqualFWR-0.3-MeasurementOutcomes}
		\end{minipage}
	\end{tcolorbox}
\end{figure}

\begin{figure}[H]
	\begin{tcolorbox}
		[
			grow to left by = 1.0 cm,
			grow to right by = 1.0 cm,
			colback = gray!03,
			enhanced jigsaw, 
			sharp corners,
			toprule = 1.0 pt,
			bottomrule = 1.0 pt,
			leftrule = 0.1 pt,
			rightrule = 0.1 pt,
			sharp corners,
			center title,
			fonttitle = \bfseries
		]
		\centering
		\begin{minipage}{0.45 \textwidth}
			\includegraphics[scale = 0.45, trim = {0 0 0cm 0}, clip]{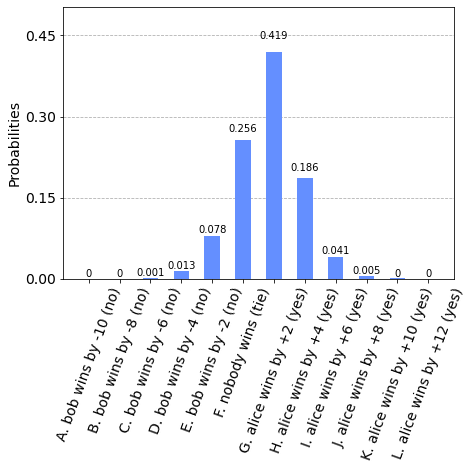}
			\caption{The possible outcomes and their probabilities when $r_A = r_B = 0.5$.}
			\label{fig:8+6+EqualFWR-0.5-MeasurementOutcomes}
		\end{minipage}
		\hfill
		\begin{minipage}{0.45 \textwidth}
			\includegraphics[scale = 0.45, trim = {0 0 0cm 0}, clip]{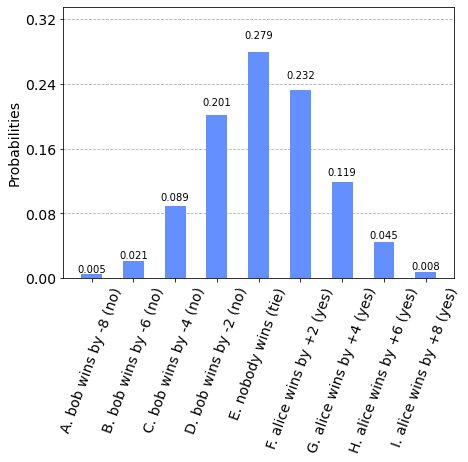}
			\caption{The possible outcomes and their probabilities when $r_A = r_B = 0.7$.}
			\label{fig:8+6+EqualFWR-0.7-MeasurementOutcomes}
		\end{minipage}
	\end{tcolorbox}
\end{figure}

Figures \ref{fig:8+6+EqualFWR-0.1-MeasurementOutcomes}--\ref{fig:8+6+EqualFWR-0.7-MeasurementOutcomes} depict intermediate scenarios, where both parties behave in a quantum manner adopting equal free will radii. The experimental data in these figures demonstrate that $p$ decreases from $1.0$, when $r_A = r_B = 0$, to a value around $0.65$ when $r_A = r_B = 0.5$, and even further to its minimum value, which is less than $0.5$, when $r_A = r_B = 1$. It is therefore evident that $p$ strictly decreases as the free will radius increases.

\textbf{Conclusion 1}. The obvious conclusion here is that \emph{the free will radius determines the degree of independence of the legislators}. As the free will radius increases, the legislators become more independent, to the point where they are completely independent when $r_A = r_B = 1$. Hence, when the free will radius tends to $1$, it is as if there are only independent representatives in the parliament. That is, all legislators behave as if they do not belong to a specific party, but, instead, have total freedom to decide how to vote.

\subsection{The parties employ different free will radii} \label{subsec:The Parties Employ Different Free Will Radii}

In the second scenario Alice and Bob's parties have again $8$ and $6$ representatives, there are no independent legislators, but now the free will radii are different, i.e., $n_A = 8$, $n_B = 6$, $n_I = 0$, and $r_A \neq r_B$.

\begin{figure}[H]
	\begin{tcolorbox}
		[
		grow to left by = 1.0 cm,
		grow to right by = 1.0 cm,
		colback = gray!03,
		enhanced jigsaw, 
		sharp corners,
		toprule = 1.0 pt,
		bottomrule = 1.0 pt,
		leftrule = 0.1 pt,
		rightrule = 0.1 pt,
		sharp corners,
		center title,
		fonttitle = \bfseries
		]
		\centering
		\begin{minipage}{0.45 \textwidth}
			\includegraphics[scale = 0.45, trim = {0 0 0cm 0}, clip]{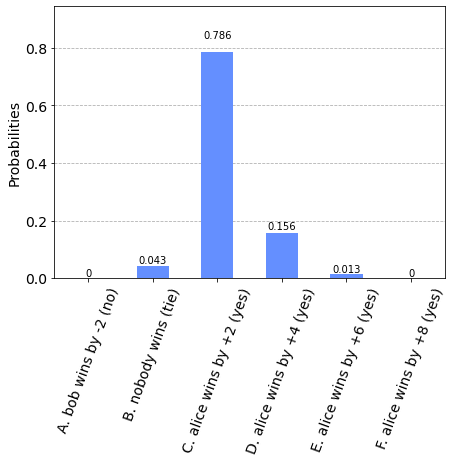}
			\caption{The possible outcomes and their probabilities when $r_A = 0.15$ and $r_B = 0.30$.}
			\label{fig:8+6+FWRA-0.15-FWRB-0.30-MeasurementOutcomes}
		\end{minipage}
		\hfill
		\begin{minipage}{0.45 \textwidth}
			\includegraphics[scale = 0.45, trim = {0 0 0cm 0}, clip]{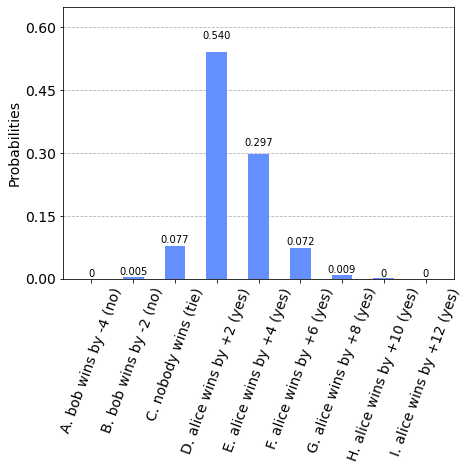}
			\caption{The possible outcomes and their probabilities when $r_A = 0.25$ and $r_B = 0.50$.}
			\label{fig:8+6+FWRA-0.25-FWRB-0.50-MeasurementOutcomes}
		\end{minipage}
	\end{tcolorbox}
\end{figure}

\begin{figure}[H]
	\begin{tcolorbox}
		[
		grow to left by = 1.0 cm,
		grow to right by = 1.0 cm,
		colback = gray!03,
		enhanced jigsaw, 
		sharp corners,
		toprule = 1.0 pt,
		bottomrule = 1.0 pt,
		leftrule = 0.1 pt,
		rightrule = 0.1 pt,
		sharp corners,
		center title,
		fonttitle = \bfseries
		]
		\centering
		\begin{minipage}{0.45 \textwidth}
			\includegraphics[scale = 0.45, trim = {0 0 0cm 0}, clip]{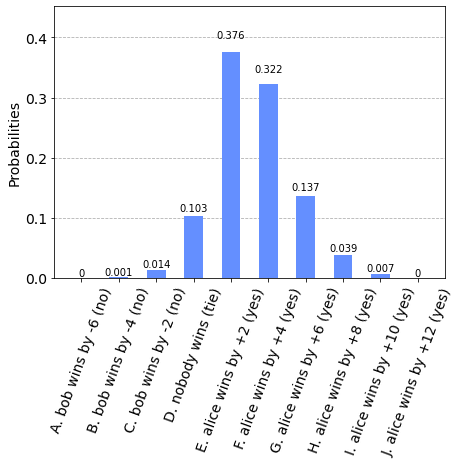}
			\caption{The possible outcomes and their probabilities when $r_A = 0.35$ and $r_B = 0.70$.}
			\label{fig:8+6+FWRA-0.35-FWRB-0.70-MeasurementOutcomes}
		\end{minipage}
		\hfill
		\begin{minipage}{0.45 \textwidth}
			\includegraphics[scale = 0.45, trim = {0 0 0cm 0}, clip]{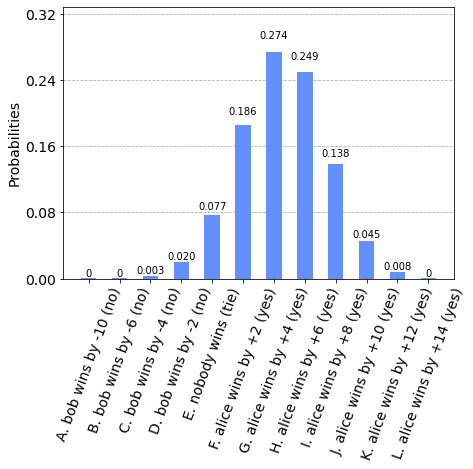}
			\caption{The possible outcomes and their probabilities when $r_A = 0.5$ and $r_B = 1$.}
			\label{fig:8+6+FWRA-0.50-FWRB-1.00-MeasurementOutcomes}
		\end{minipage}
	\end{tcolorbox}
\end{figure}

If the minority party uses a greater free will radius than the majority party, $r_A < r_B$, then the probability $p$ that the bill passes is visibly increased compared to the case where both parties have the same free will radii. This is clearly demonstrated by comparing the experimental data contained in Figures \ref{fig:8+6+FWRA-0.15-FWRB-0.30-MeasurementOutcomes}--\ref{fig:8+6+FWRA-0.50-FWRB-1.00-MeasurementOutcomes} with those in Figures \ref{fig:8+6+EqualFWR-0.1-MeasurementOutcomes}--\ref{fig:8+6+EqualFWR-0.7-MeasurementOutcomes}.

\begin{figure}[H]
	\begin{tcolorbox}
		[
		grow to left by = 1.0 cm,
		grow to right by = 1.0 cm,
		colback = gray!03,
		enhanced jigsaw, 
		sharp corners,
		toprule = 1.0 pt,
		bottomrule = 1.0 pt,
		leftrule = 0.1 pt,
		rightrule = 0.1 pt,
		sharp corners,
		center title,
		fonttitle = \bfseries
		]
		\centering
		\begin{minipage}{0.45 \textwidth}
			\includegraphics[scale = 0.45, trim = {0 0 0cm 0}, clip]{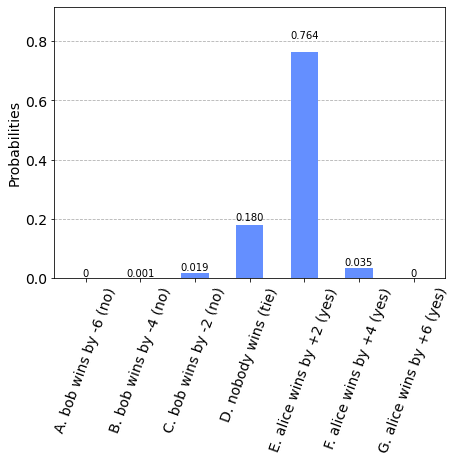}
			\caption{The possible outcomes and their probabilities when $r_A = 0.30$ and $r_B = 0.15$.}
			\label{fig:8+6+FWRA-0.30-FWRB-0.15-MeasurementOutcomes}
		\end{minipage}
		\hfill
		\begin{minipage}{0.45 \textwidth}
			\includegraphics[scale = 0.45, trim = {0 0 0cm 0}, clip]{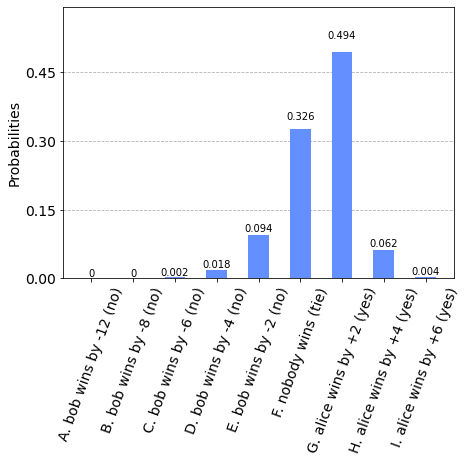}
			\caption{The possible outcomes and their probabilities when $r_A = 0.50$ and $r_B = 0.25$.}
			\label{fig:8+6+FWRA-0.50-FWRB-0.25-MeasurementOutcomes}
		\end{minipage}
	\end{tcolorbox}
\end{figure}

The tests shown in Figures \ref{fig:8+6+FWRA-0.15-FWRB-0.30-MeasurementOutcomes}--\ref{fig:8+6+FWRA-0.50-FWRB-1.00-MeasurementOutcomes} were conducted assuming the free will radius of Bob's party was always twice that of Alice's party, i.e., $r_B = 2 r_A$. Such a scenario favors Alice's party, which obtains an overwhelming advantage to pass the bill, with the probability $p \approx 0.9$ in most cases.

\begin{figure}[H]
	\begin{tcolorbox}
		[
		grow to left by = 1.0 cm,
		grow to right by = 1.0 cm,
		colback = gray!03,
		enhanced jigsaw, 
		sharp corners,
		toprule = 1.0 pt,
		bottomrule = 1.0 pt,
		leftrule = 0.1 pt,
		rightrule = 0.1 pt,
		sharp corners,
		center title,
		fonttitle = \bfseries
		]
		\centering
		\begin{minipage}{0.45 \textwidth}
			\includegraphics[scale = 0.45, trim = {0 0 0cm 0}, clip]{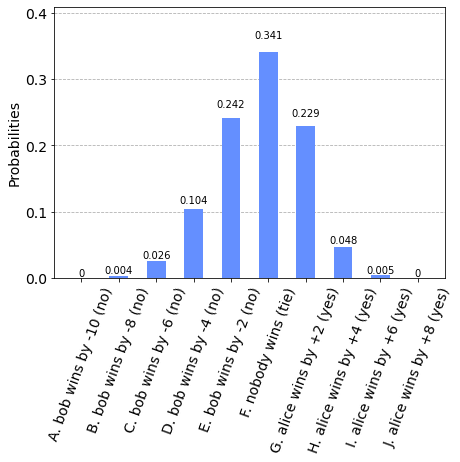}
			\caption{The possible outcomes and their probabilities when $r_A = 0.70$ and $r_B = 0.35$.}
			\label{fig:8+6+FWRA-0.70-FWRB-0.35-MeasurementOutcomes}
		\end{minipage}
		\hfill
		\begin{minipage}{0.45 \textwidth}
			\includegraphics[scale = 0.45, trim = {0 0 0cm 0}, clip]{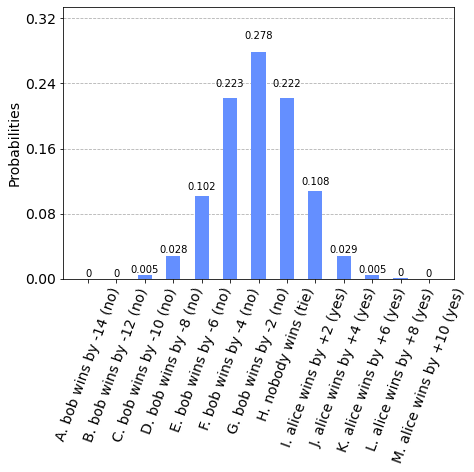}
			\caption{The possible outcomes and their probabilities when $r_A = 1.0$ and $r_B = 0.5$.}
			\label{fig:8+6+FWRA-1.00-FWRB-0.50-MeasurementOutcomes}
		\end{minipage}
	\end{tcolorbox}
\end{figure}

Things get much more interesting in the opposite direction, when the majority party employs a greater free will radius than the minority party: $r_B < r_A$. Then, the probability $p$ that the bill passes is visibly decreased compared to the case where both parties have the same free will radii. This conclusion can be deduced from the experimental data contained in Figures \ref{fig:8+6+FWRA-0.30-FWRB-0.15-MeasurementOutcomes}--\ref{fig:8+6+FWRA-1.00-FWRB-0.50-MeasurementOutcomes}. The results obtained in these figures were obtained assuming the free will radius of Alice's party was always twice that of Bob's party, i.e., $r_A = 2 r_B$. In contrast to the previous case, this scenario favors Bob's party, which now stands a very good chance to block the bill. As the figures indicate, if Bob's free will radius is half of Alice's, and the latter surpasses $0.5$, the probability $p$ drops below $0.5$, meaning that it is more probable now for the bill to fail than to pass.

\textbf{Conclusion 2}. The decisive role of the free will radius, or, equivalently, of the degree of independence, is confirmed by the experiments of this subsection. If the free will radii are not equal, then the odds improve for the party that employs the smaller free will radius. The party that grants the greater freedom to its representatives is more prone to lose the vote, and the greater this freedom is, the greater is the chance of this happening.

\subsection{The effect of independent legislators} \label{subsec:The Effect of Independent Legislators}

Initially, we assume that Alice's party holds the majority with $8$ representatives, Bob's party has $4$ representatives, both parties employ equal free will radii, but now there are also $2$ independent legislators, i.e., $n_A = 8$, $n_B = 4$, $n_I = 2$, and $r_A = r_B$. To understand how the presence of independent legislators affects the operation of the quantum parliament, we begin by examining the two extreme cases.

\begin{figure}[H]
	\begin{tcolorbox}
		[
		grow to left by = 1.0 cm,
		grow to right by = 1.0 cm,
		colback = gray!03,
		enhanced jigsaw, 
		sharp corners,
		toprule = 1.0 pt,
		bottomrule = 1.0 pt,
		leftrule = 0.1 pt,
		rightrule = 0.1 pt,
		sharp corners,
		center title,
		fonttitle = \bfseries
		]
		\centering
		\begin{minipage}{0.45 \textwidth}
			\includegraphics[scale = 0.45, trim = {0 0 0cm 0}, clip]{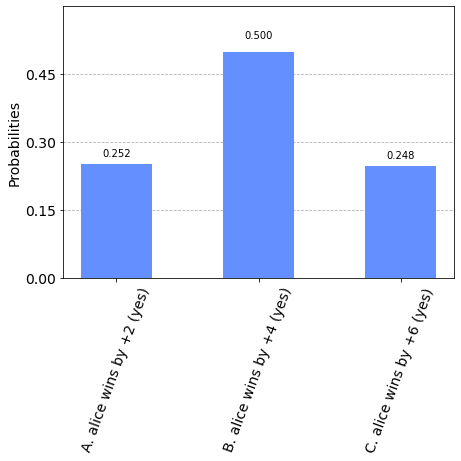}
			\caption{The possible outcomes and their probabilities with $2$ independent legislators and $r_A = r_B = 0$.}
			\label{fig:8+4+2+EqualFWR-0.0-MeasurementOutcomes}
		\end{minipage}
		\hfill
		\begin{minipage}{0.45 \textwidth}
			\includegraphics[scale = 0.45, trim = {0 0 0cm 0}, clip]{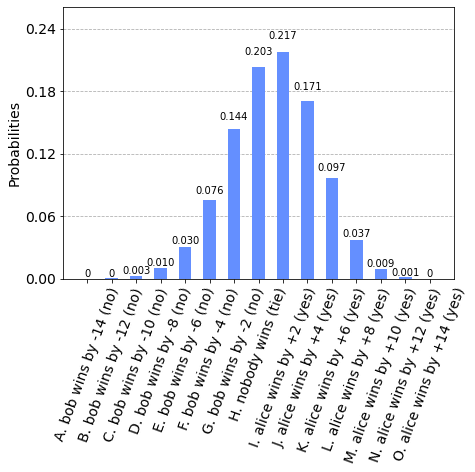}
			\caption{The possible outcomes and their probabilities with $2$ independent legislators and $r_A = r_B = 1$.}
			\label{fig:8+4+2+EqualFWR-1.0-MeasurementOutcomes}
		\end{minipage}
	\end{tcolorbox}
\end{figure}

\begin{itemize}
	\item	This time, even if both parties set their free will radii to zero, $r_A = r_B = 0$, the parliament does not become completely classical, in the sense that the voting outcome is no longer unique. It is still predetermined that the bill will pass, i.e., $p = 1.0$, but what is not certain a priori is how many representatives will vote in favor of the bill. The probability distribution of the outcomes is shown in Figure \ref{fig:8+4+2+EqualFWR-0.0-MeasurementOutcomes}. Now the most probable outcome is that the bill will pass with a margin of $4$ votes instead of $2$. Hence, the presence of independent legislators in the event of zero free will radii, will simply result to a wider margin of surplus votes for the bill. Let us remark that this situation is experimentally distinguishable from the analogous situation in subsection \ref{subsec:The Free Will Radius determines Independence}, as can be seen by a direct comparison of Figure \ref{fig:8+4+2+EqualFWR-0.0-MeasurementOutcomes} with Figure \ref{fig:8+6+EqualFWR-0.0-MeasurementOutcomes}.
	\item	The other extreme scenario is when both parties allow complete freedom of choice to their representatives, by setting their free will radii to the maximum value: $r_A = r_B = 1$. As shown in Figure \ref{fig:8+4+2+EqualFWR-1.0-MeasurementOutcomes}, the voting outcome can be anything. As explained before, when the free will radius becomes $1$, the party representatives become independent. Therefore, this parliament behaves as if it consists entirely of independent legislators and is statistically identical with the one depicted in Figure \ref{fig:8+6+EqualFWR-1.0-MeasurementOutcomes}.
\end{itemize}

\begin{figure}[H]
	\begin{tcolorbox}
		[
		grow to left by = 1.0 cm,
		grow to right by = 1.0 cm,
		colback = gray!03,
		enhanced jigsaw, 
		sharp corners,
		toprule = 1.0 pt,
		bottomrule = 1.0 pt,
		leftrule = 0.1 pt,
		rightrule = 0.1 pt,
		sharp corners,
		center title,
		fonttitle = \bfseries
		]
		\centering
		\begin{minipage}{0.45 \textwidth}
			\includegraphics[scale = 0.45, trim = {0 0 0cm 0}, clip]{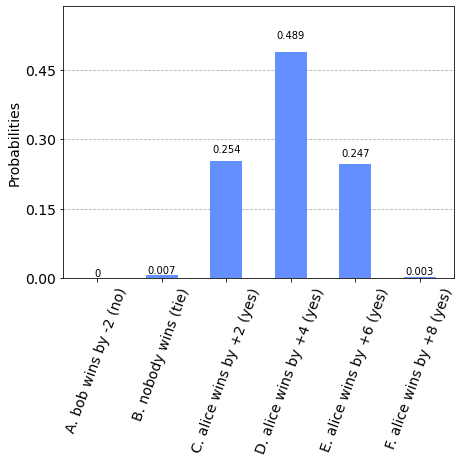}
			\caption{The possible outcomes and their probabilities with $2$ independent legislators and $r_A = r_B = 0.1$.}
			\label{fig:8+4+2+EqualFWR-0.1-MeasurementOutcomes}
		\end{minipage}
		\hfill
		\begin{minipage}{0.45 \textwidth}
			\includegraphics[scale = 0.45, trim = {0 0 0cm 0}, clip]{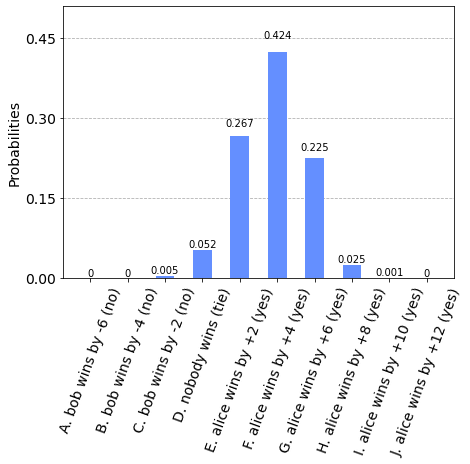}
			\caption{The possible outcomes and their probabilities with $2$ independent legislators and $r_A = r_B = 0.3$.}
			\label{fig:8+4+2+EqualFWR-0.3-MeasurementOutcomes}
		\end{minipage}
	\end{tcolorbox}
\end{figure}

\begin{figure}[H]
	\begin{tcolorbox}
		[
		grow to left by = 1.0 cm,
		grow to right by = 1.0 cm,
		colback = gray!03,
		enhanced jigsaw, 
		sharp corners,
		toprule = 1.0 pt,
		bottomrule = 1.0 pt,
		leftrule = 0.1 pt,
		rightrule = 0.1 pt,
		sharp corners,
		center title,
		fonttitle = \bfseries
		]
		\centering
		\begin{minipage}{0.45 \textwidth}
			\includegraphics[scale = 0.45, trim = {0 0 0cm 0}, clip]{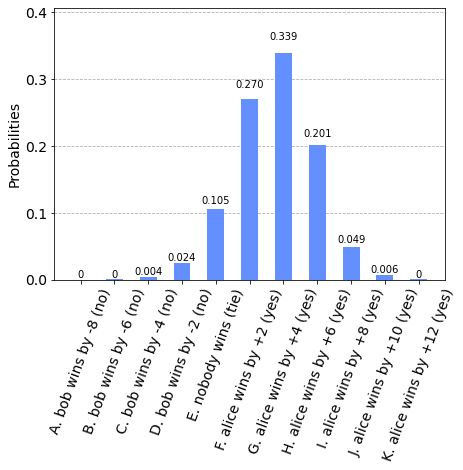}
			\caption{The possible outcomes and their probabilities with $2$ independent legislators and $r_A = r_B = 0.5$.}
			\label{fig:8+4+2+EqualFWR-0.5-MeasurementOutcomes}
		\end{minipage}
		\hfill
		\begin{minipage}{0.45 \textwidth}
			\includegraphics[scale = 0.45, trim = {0 0 0cm 0}, clip]{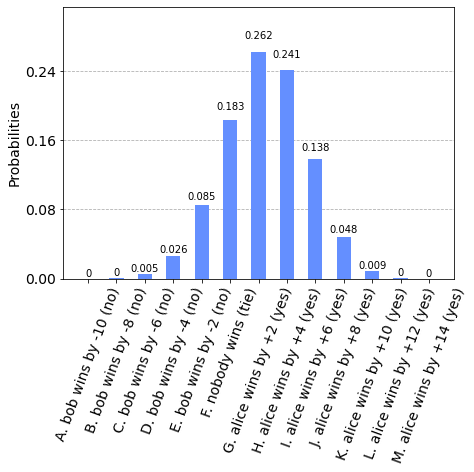}
			\caption{The possible outcomes and their probabilities with $2$ independent legislators and $r_A = r_B = 0.7$.}
			\label{fig:8+4+2+EqualFWR-0.7-MeasurementOutcomes}
		\end{minipage}
	\end{tcolorbox}
\end{figure}

Some intermediate scenarios are visualized in Figures \ref{fig:8+4+2+EqualFWR-0.1-MeasurementOutcomes}--\ref{fig:8+4+2+EqualFWR-0.7-MeasurementOutcomes}. These aim to demonstrate the effect of the independent legislators on the voting process as a function of the free will radii adopted by the parties. A simple comparison of the Figures \ref{fig:8+4+2+EqualFWR-0.1-MeasurementOutcomes}--\ref{fig:8+4+2+EqualFWR-0.7-MeasurementOutcomes} with the Figures \ref{fig:8+6+EqualFWR-0.1-MeasurementOutcomes}--\ref{fig:8+6+EqualFWR-0.7-MeasurementOutcomes} shows that the presence of independent legislators has two detectable effects.

\begin{enumerate}
	\item	They increase the probability $p$ that the bill will pass.
	\item 	They produce \emph{new} outcomes because it is now probable that the bill will gather more votes in favor, than before.
\end{enumerate}

Of course, as the free will radius tends to $1$, both of these effects fade to the point of becoming unobservable, which is in line with the fact that the increase of the free will radius makes the representatives more independent.

\textbf{Conclusion 3}. The experimental evidence presented above shows that the existence of independent representatives causes more subtle results than a variation in the free will radius. For small values of the free will radii they act as potential ally of the majority party. For higher values of the free will radii their presence is less discernible. The fact is that in this scenario the operation of the parliament is quite similar to the one encountered in subsection \ref{subsec:The Free Will Radius determines Independence}. Thus, this begs the question whether the role of the independent representatives is significant or necessary at all.

To answer this question we contrast two new scenarios. In both of these scenarios Bob's party has exactly as many legislators as Alice's party and both parties adopt equal free will radii. Their only difference is that in the first scenario there are no independent legislators, i.e., $n_A = n_B = 8$, $n_I = 0$, and $r_A = r_B$, whereas in the latter there are $2$ independent legislators, i.e., $n_A = n_B = 6$, $n_I = 2$, and $r_A = r_B$. It turns out that the role of the independent legislators is important in situations where both parties have the same number of representatives. This is easily seen by comparing the three sets of figures shown below.

\begin{figure}[H]
	\begin{tcolorbox}
		[
		grow to left by = 1.0 cm,
		grow to right by = 1.0 cm,
		colback = gray!03,
		enhanced jigsaw, 
		sharp corners,
		toprule = 1.0 pt,
		bottomrule = 1.0 pt,
		leftrule = 0.1 pt,
		rightrule = 0.1 pt,
		sharp corners,
		center title,
		fonttitle = \bfseries
		]
		\centering
		\begin{minipage}[b]{0.45 \textwidth}
			\includegraphics[scale = 0.45, trim = {0 0 0cm 0}, clip]{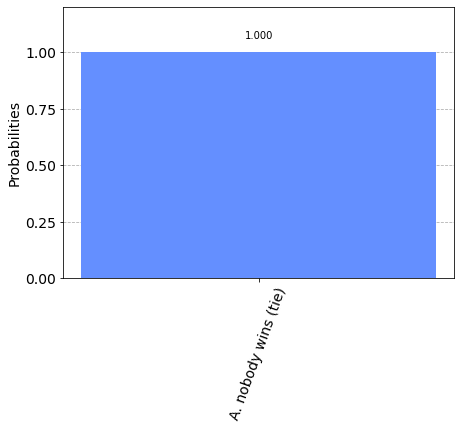}
			\caption{The unique possible outcome when $n_A = n_B = 8$, $r_A = r_B = 0$, and $n_I = 0$.}
			\label{fig:7+7+EqualFWR-0.0-MeasurementOutcomes}
		\end{minipage}
		\hfill
		\begin{minipage}[b]{0.45 \textwidth}
			\includegraphics[scale = 0.45, trim = {0 0 0cm 0}, clip]{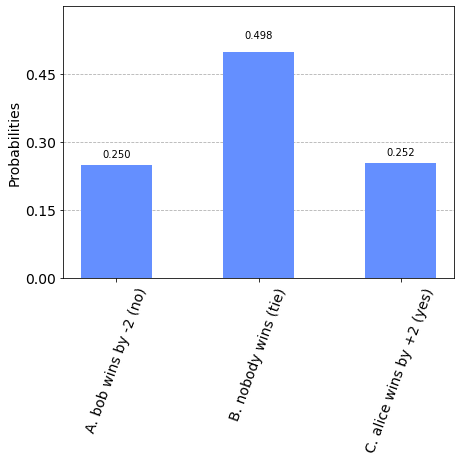}
			\caption{The three possible outcomes when $n_A = n_B = 6$, $r_A = r_B = 0$, and $n_I = 0$.}
			\label{fig:6+6+2+EqualFWR-0.0-MeasurementOutcomes}
		\end{minipage}
	\end{tcolorbox}
\end{figure}

\begin{figure}[H]
	\begin{tcolorbox}
		[
		grow to left by = 1.0 cm,
		grow to right by = 1.0 cm,
		colback = gray!03,
		enhanced jigsaw, 
		sharp corners,
		toprule = 1.0 pt,
		bottomrule = 1.0 pt,
		leftrule = 0.1 pt,
		rightrule = 0.1 pt,
		sharp corners,
		center title,
		fonttitle = \bfseries
		]
		\centering
		\begin{minipage}{0.45 \textwidth}
			\includegraphics[scale = 0.45, trim = {0 0 0cm 0}, clip]{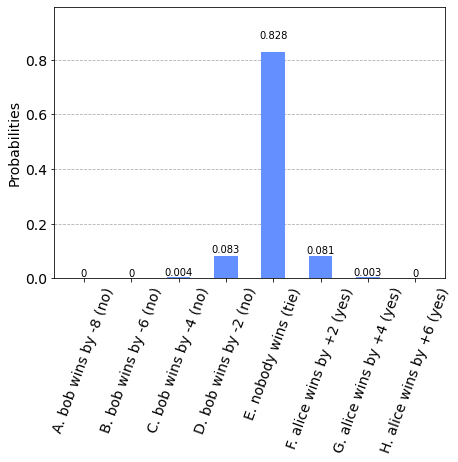}
			\caption{The possible outcomes when $n_A = n_B = 8$, $r_A = r_B = 0.2$, and $n_I = 0$.}
			\label{fig:7+7+EqualFWR-0.2-MeasurementOutcomes}
		\end{minipage}
		\hfill
		\begin{minipage}{0.45 \textwidth}
			\includegraphics[scale = 0.45, trim = {0 0 0cm 0}, clip]{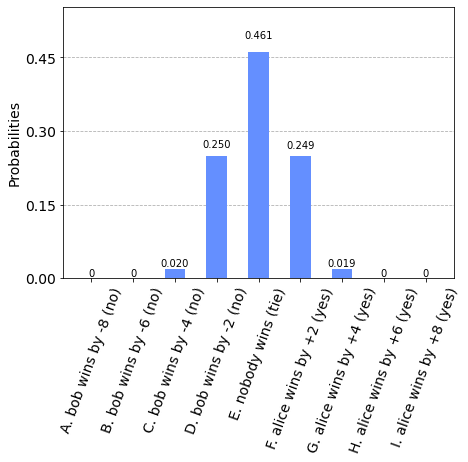}
			\caption{The possible outcomes when $n_A = n_B = 6$, $r_A = r_B = 0.2$, and $n_I = 2$.}
			\label{fig:6+6+2+EqualFWR-0.2-MeasurementOutcomes}
		\end{minipage}
	\end{tcolorbox}
\end{figure}

\begin{figure}[H]
	\begin{tcolorbox}
		[
		grow to left by = 1.0 cm,
		grow to right by = 1.0 cm,
		colback = gray!03,
		enhanced jigsaw, 
		sharp corners,
		toprule = 1.0 pt,
		bottomrule = 1.0 pt,
		leftrule = 0.1 pt,
		rightrule = 0.1 pt,
		sharp corners,
		center title,
		fonttitle = \bfseries
		]
		\centering
		\begin{minipage}{0.45 \textwidth}
			\includegraphics[scale = 0.45, trim = {0 0 0cm 0}, clip]{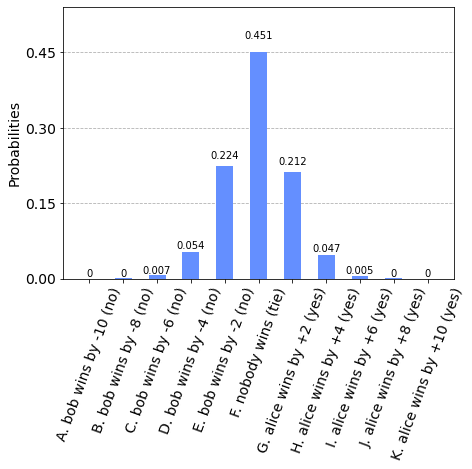}
			\caption{The possible outcomes when $n_A = n_B = 8$, $r_A = r_B = 0.5$, and $n_I = 0$.}
			\label{fig:7+7+EqualFWR-0.5-MeasurementOutcomes}
		\end{minipage}
		\hfill
		\begin{minipage}{0.45 \textwidth}
			\includegraphics[scale = 0.45, trim = {0 0 0cm 0}, clip]{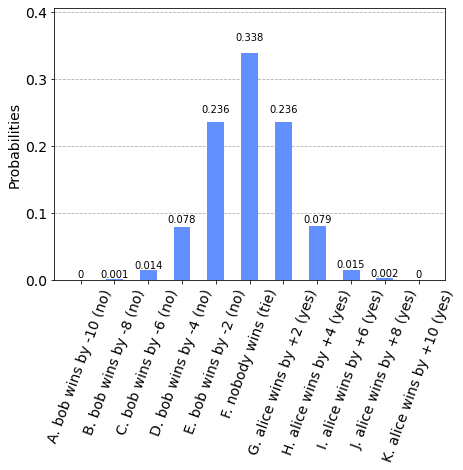}
			\caption{The possible outcomes when $n_A = n_B = 6$, $r_A = r_B = 0.5$, and $n_I = 2$.}
			\label{fig:6+6+2+EqualFWR-0.5-MeasurementOutcomes}
		\end{minipage}
	\end{tcolorbox}
\end{figure}

If both free will radii are zero, then when there are no independent legislators the probability $p$ that the bill will pass is precisely $0.0$ as shown in Figure \ref{fig:7+7+EqualFWR-0.0-MeasurementOutcomes}. The existence of $2$ independent representatives changes the situation significantly (see Figure \ref{fig:6+6+2+EqualFWR-0.0-MeasurementOutcomes}), and increases the probability $p$ to $0.25$. If the free will radii become $0.2$, then when there are no independent legislators the probability $p$ that the bill will pass is less than $0.1$, according to Figure \ref{fig:7+7+EqualFWR-0.2-MeasurementOutcomes}. The effect of the presence of the $2$ independent representatives, depicted in Figure \ref{fig:6+6+2+EqualFWR-0.2-MeasurementOutcomes}, is to increase again the probability $p$ to around $0.26-0.27$. If the free will radii increase further, let's say up to $0.5$, then even when there are no independent legislators, the probability $p$ that the bill will pass goes up to slightly above $0.25$, as can be seen in Figure \ref{fig:7+7+EqualFWR-0.5-MeasurementOutcomes}. The effect of the $2$ independent representatives, as visualized in Figure \ref{fig:6+6+2+EqualFWR-0.5-MeasurementOutcomes}, is to provide a significantly lower increase to the probability $p$, driving it to around $0.33$.

\textbf{Conclusion 4}. The experimental data indicate that the independent legislators play an important role when synthesis of the parliament may lead to a political deadlock. When no party has the numerical advantage and it is certain that the bill will not pass, the presence of the independent representatives may provide a significant thrust to the voting procedure, by turning an impossibility to a probable eventuality. Such a situation can only happen when the parties are almost classical in the sense that they employ zero or very small free will radii. When the free will radii increase (say over $0.5$), the effect of the independent legislators becomes negligible. Therefore, in a classical parliament where two parties have the same number of representatives their presence is useful and may even deemed necessary. In a true quantum parliament their value is diminished.

\section{The game of ``Passing the Bill''} \label{sec:The Game of Passing the Bill}

In such virtual quantum parliament we may envision a repeated game where Alice's party, which proposes and tries to pass the bill, is considered player 1, and Bob's party, which is strongly against the bill and tries to block it, is player 2. In that way, for each bill both players play a stage of the game ``Passing the Bill.''

Let us point out that quantum games is a noteworthy and dynamic field of the quantum era. Quantum games have addressed important and, sometimes, difficult problems in the quantum realm. We present the voting process as a quantum game in an attempt to make the whole setting more mnemonic and playful. One may recall that cryptographic protocols are usually presented as games between the two fictional heroes Alice and Bob, who are the two remote parties trying to exchange messages, and also involving Eve who tries to eavesdrop the conversation. A typical and convincing example is the quantum game of coin tossing and its relation to the BB84 protocol (see \cite{Bennett1984}, \cite{Bennett2014} and the more recent \cite{Ampatzis2021}). This setting has already been generalized to higher dimensions in \cite{Aharon2010}. For the interested reader we mention two landmark papers in the field of quantum game theory going back to 1999, which helped shape the field. One is Meyer's PQ penny flip game \cite{Meyer1999}, which can be regarded as the quantum analogue of the classical penny flip game, and other is the introduction of the Eisert-Wilkens-Lewenstein scheme \cite{Eisert1999} that is widely used in this area. With respect to the PQ penny flip game, some recent results are presented in \cite{Andronikos2018} and \cite{Andronikos2021}. The Eisert-Wilkens-Lewenstein scheme was especially successful in quantum adaptations of the famous Prisoners' Dilemma, where the quantum strategies outperformed the classical strategy (\cite{Eisert1999}), and also to quantum extensions of the classical repeated Prisoners' Dilemma conditional strategies (\cite{Giannakis2019}). Winning strategies for abstract quantum games can also be encoded as infinite words accepted by quantum automata, as demonstrated in \cite{Giannakis2015a}. Closing this brief discussion, we note that the idea to resort to unconventional means to achieve better than classical results in games is not limited to the quantum domain. The most iconic classical games, such as the Prisoners' Dilemma, have also been cast in terms of biological notions (see \cite{Kastampolidou2020}, \cite{Kastampolidou2021}, \cite{Papalitsas2021}, and \cite{Kastampolidou2022} for further details and more references). It turns out that a lot of classical games can be expressed in the context of biological and bio-inspired processes (see \cite{Theocharopoulou2019, Kastampolidou2020a} for more references), and this metaphor can even be applied to computer viruses (see \cite{Kostadimas2021} and references therein).

In the case of a quantum parliament, the resulting ``Passing the Bill'' game is somewhat complex because neither Alice nor Bob know a priory whether they will be successful in their efforts to pass or stop, respectively, the bill in question. Hence, it can be modeled as a game that incorporates chance moves.
Games of such type are typically analyzed by probabilistic means (see for in depth treatment the very accessible  books \cite{Maschler2020}, \cite{Dixit2015} or \cite{Tadelis2013}). To make the analysis of the game more tractable and easier to follow, we shall assume that both Alice and Bob use the \emph{same} free will radius, and, that it is conceivable that one or both may, in exceptional circumstances, cancel this privilege of diversification and demand total obedience from their representatives.

The rationale behind this game is the following. Both Alice and Bob are faced with an important choice regarding their current situation. They do not know a priory with absolute certainty, that is with probability $1.0$, what the outcome of the voting process will be. They can only draw from their previous experience to estimate probabilistically what are the odds that their adopted strategy will succeed in the current situation. So when playing the game of ``Passing the Bill,'' both parties must make a critical decision.

\begin{itemize}
	\item	Should they follow their normal practice and tolerate the fact that their representatives may vote withing the free will radius allowed by the party, or
	\item	should temporarily abolish entirely the free will radius, demanding that their representatives exercise utmost discipline and unwavering commitment to the decision made by leadership of the party.
\end{itemize}

The normal behavior that encourages and accepts diversification withing the free will radius constitutes the \emph{tolerant} strategy. The opposite strategy, which demands obedience to the party decision, is called the \emph{autocratic} strategy. In modern democracies one of the hallmark characteristics of the political establishment is the trend for greater freedom of expression and less autocratic behavior on behalf of all political agents. Hence, while the tolerant strategy is welcomed by the political institutions and the society as a whole, the autocratic strategy is frowned upon, and induces a certain \emph{political cost} to the party that resorts to such means, e.g., fewer seats in the next parliament. Nonetheless, according to the accumulated previous experience, as corroborated by the extensive experimental tests of section \ref{sec:Simulating the Quantum Parliament}, the autocratic behavior maximizes the chances of success for each party in the game of ``Passing the Bill.'' Therefore, in certain rare cases where the bill in question is deemed of particular importance, it may be a viable strategy for either or both the parties. So, the game may unfold according to following premises.

\begin{itemize}
	\item[\textbf{H1:}]	When using the \emph{tolerant} strategy, both parties receive the \emph{same reward} $r > 0$, representing their political gain, as payoff when they succeed in passing or stopping a bill, and $0$ payoff when they fail to do so.
	\item[\textbf{H2:}]	When using the \emph{autocratic} strategy, both parties spend the \emph{same} amount of \emph{political capital} because they adopt a strategy that is disapproved by the society. This loss of political capital is modeled by a constant cost $c > 0$ that is subcontracted from the payoff they would otherwise receive, had they used the \emph{tolerant} strategy. Specifically, instead of the parties receiving the typical reward $r$ and $0$ payoffs, they will receive the reduced payoffs $r - c$ and $- c$ in case of success and failure, respectively. It is assumed that $c < r$, so that the parties can still hope for a positive, although diminished, payoff in case of success.
\end{itemize}

According to Alice and Bob's accumulated past experience in previous games, they both ascribe the same probability $p$ to the event that the bill will pass and $1 - p$ to the event that the bill will fail to pass, in case they both employ the \emph{same strategy} irrespective of whether this is the tolerant or the autocratic strategy. If one party sticks to the tolerant strategy, but the other party switches to the autocratic strategy, the latter increases the probability of success (in either passing or stopping the bill) by a positive constant $0 < \varepsilon < 1$. As demonstrated experimentally in section \ref{sec:Simulating the Quantum Parliament}, the probability $p$ and the increment $\varepsilon$ will, in general, vary depending on the precise synthesis of the parliament, the exact values of the free will radii, and the presence or not of independent legislators. Undoubtedly, the constant $c$ representing the political cost incurred because of the use of the autocratic strategy plays a pivotal role in this game. If $c$ is small relative to the reward $r$, then it will tempting for both parties to resort to the autocratic strategy, something that will, necessarily, degrade the quantum parliament to a classical one. If, on the other hand, $c$ is high enough relative to the reward $r$, so that the use of the autocratic strategy is prohibitive, then the parliament will remain quantum. Ultimately, the parameter $c$ will depend on how much the political system and the society as a whole value pluralism and freedom of choice.


\section{Discussion and conclusions} \label{sec:Discussion & Conclusions}

In this paper we have presented the first, to the best of our knowledge, functional model of a quantum parliament. The parliament is comprised primarily of two parties, as well as a number of independent legislators. In our theoretical analysis of the proposed model we identified a single critical parameter, namely the free will radius, that can be used as a practical measure of the quantumness of the parties and the parliament as a whole. The free will radius used by the two parties determines the degree of independence that is afforded to the representatives of the parties. Setting the free will radius to zero degrades the quantum parliament to a classical one, whereas setting the free will radius to its maximum value $1$, makes the representatives totally independent. This observation has motivated us to propose the game of ``Passing the Bill,'' which captures the operation of the quantum parliament and the dilemmata posed to the leadership of the two parties.

We have also constructed a quantum circuit with which we were able to simulate the operation of the quantum parliament under various scenarios. Extensive experimental tests have led us to the following insightful conclusions.

\begin{itemize}
	\item	The degree of independence of the representatives is strictly increasing with the free will radius. As the free will radius increases, the legislators become more independent, to the point where they are completely independent when the free will radius takes its maximum value $1$. In such a case the legislators behave as if they do not belong to a specific party, but, instead, have total freedom to decide how to vote. One could go as far as to say that free will maters more than numbers, in the sense that the majority party may still lose a vote, if its representatives are allowed enough freedom of choice.
	\item	The critical role of the free will radius becomes more evident when the two parties employ different free will radii. In such a case, then the odds improve drastically for the party that employs the smaller free will radius. The party that grants the greater freedom to its representatives is more prone to lose the vote.
	\item	The existence or not of independent representatives in a quantum parliament is of secondary importance, compared to the free will radius. For small values of the free will radii they act as potential ally of the majority party. For higher values of the free will radii their presence is less discernible.
	\item	The presence of independent legislators is important only when no party has the numerical advantage, in which case they may provide a significant impetus to the voting procedure, by turning an impossibility to a probable eventuality. Note though that such a situation can only happen when the parties employ zero or very small free will radii. When the free will radii increase (say over $0.5$), the effect of the independent legislators becomes negligible. Therefore, in an almost classical parliament where two parties have the same number of representatives their presence is useful and may even deemed necessary. In a true quantum parliament their value is diminished.
\end{itemize}

At the end of the day, the bottom line in accessing the usefulness of a political process is the resulting social welfare. In this paper we studied whether the shift to the quantum paradigm offers any discernible improvement over the current classical institutions, and, if any, under what conditions exactly. In closing let us say that many more directions can and should be investigated. For instance, in a quantum setting, should all the entities involved, i.e., voters, parties, politicians, bills, etc. be purely classical and just resort to quantum means for the voting process, or are they themselves quantum? This is an fundamental question of a rather philosophical nature, that is probably very hard, to answer and, in our view, it deserves further consideration.

\bibliographystyle{ieeetr}
\bibliography{ATwo-PartyQuantumParliament}

\begin{thebibliography}{10}

\bibitem{Aerts2005}
D.~Aerts, ``Towards a new democracy: Consensus through quantum parliament,'' in
  {\em Worldviews, Science and Us}, World Scientific, 2005.

\bibitem{Christandl2005}
M.~Christandl and S.~Wehner, ``Quantum anonymous transmissions,'' in {\em
  Lecture Notes in Computer Science}, pp.~217--235, Springer Berlin Heidelberg,
  2005.

\bibitem{Hillery2006}
M.~Hillery, M.~Ziman, V.~Bu{\v{z}}ek, and M.~Bielikov{\'{a}}, ``Towards
  quantum-based privacy and voting,'' {\em Physics Letters A}, vol.~349,
  no.~1-4, pp.~75--81, 2006.

\bibitem{Vaccaro2007}
J.~A. Vaccaro, J.~Spring, and A.~Chefles, ``Quantum protocols for anonymous
  voting and surveying,'' {\em Physical Review A}, vol.~75, no.~1, p.~012333,
  2007.

\bibitem{Li2008}
Y.~Li and G.~Zeng, ``Quantum anonymous voting systems based on entangled
  state,'' {\em Optical Review}, vol.~15, no.~5, pp.~219--223, 2008.

\bibitem{Wang2013}
W.~Yu-Wu, W.~Xiang-He, and Z.~Zhao-Hui, ``Quantum voting protocols based on the
  non-symmetric quantum channel with controlled quantum operation
  teleportation,'' {\em Acta Physica Sinica}, vol.~62, no.~16, p.~160302, 2013.

\bibitem{Tian2015}
J.-H. Tian, J.-Z. Zhang, and Y.-P. Li, ``A voting protocol based on the
  controlled quantum operation teleportation,'' {\em International Journal of
  Theoretical Physics}, vol.~55, no.~5, pp.~2303--2310, 2015.

\bibitem{Zhang2017a}
J.-L. Zhang, S.-C. Xie, and J.-Z. Zhang, ``An elaborate secure quantum voting
  scheme,'' {\em International Journal of Theoretical Physics}, vol.~56,
  no.~10, pp.~3019--3028, 2017.

\bibitem{Xue2017}
P.~Xue and X.~Zhang, ``A simple quantum voting scheme with multi-qubit
  entanglement,'' {\em Scientific Reports}, vol.~7, no.~1, 2017.

\bibitem{Sun2018}
X.~Sun, Q.~Wang, P.~Kulicki, and M.~Sopek, ``A simple voting protocol on
  quantum blockchain,'' {\em International Journal of Theoretical Physics},
  vol.~58, no.~1, pp.~275--281, 2018.

\bibitem{Wang2019}
S.~lan Wang, S.~Zhang, Q.~Wang, and R.~hua Shi, ``Fault-tolerant quantum
  anonymous voting protocol,'' {\em International Journal of Theoretical
  Physics}, vol.~58, no.~3, pp.~1008--1016, 2019.

\bibitem{Zhang2019}
S.~Zhang, S.~lan Wang, Q.~Wang, and R.~hua Shi, ``Quantum anonymous voting
  protocol with the privacy protection of the candidate,'' {\em International
  Journal of Theoretical Physics}, vol.~58, no.~10, pp.~3323--3332, 2019.

\bibitem{Gao2021}
W.~Gao and L.~Yang, ``Quantum election protocol based on quantum public key
  cryptosystem,'' {\em Security and Communication Networks}, vol.~2021,
  pp.~1--15, 2021.

\bibitem{Arapinis2021}
M.~Arapinis, N.~Lamprou, E.~Kashefi, and A.~Pappa, ``Definitions and security
  of quantum electronic voting,'' {\em {ACM} Transactions on Quantum
  Computing}, vol.~2, no.~1, pp.~1--33, 2021.

\bibitem{Pluchino2011}
A.~Pluchino, C.~Garofalo, A.~Rapisarda, S.~Spagano, and M.~Caserta,
  ``Accidental politicians: How randomly selected legislators can improve
  parliament efficiency,'' {\em Physica A: Statistical Mechanics and its
  Applications}, vol.~390, no.~21-22, pp.~3944--3954, 2011.

\bibitem{Nielsen2010}
M.~A. Nielsen and I.~L. Chuang, {\em Quantum computation and quantum
  information}.
\newblock Cambridge University Press, 2010.

\bibitem{Wilde2018}
M.~M. Wilde, {\em Quantum Information Theory}.
\newblock Cambridge University Press, 2018.

\bibitem{Qiskit2022}
Qiskit, ``Qiskit open-source quantum development.'' \url{https://qiskit.org}.
\newblock Accessed: 2021-12-22.

\bibitem{Williams2010}
C.~Williams, {\em Explorations in Quantum Computing}.
\newblock Texts in Computer Science, Springer London, 2010.

\bibitem{Kaye2007}
P.~Kaye, R.~Laflamme, and M.~Mosca, {\em An Introduction to Quantum Computing}.
\newblock OUP Oxford, 2007.

\bibitem{QiskitUGate2022}
Qiskit, ``Ugate.''
  \url{https://qiskit.org/documentation/stubs/qiskit.circuit.library.UGate.html}.
\newblock Accessed: 2022-01-03.

\bibitem{DraperQFTAdder2022}
Qiskit, ``Draperqftadder.''
  \url{https://qiskit.org/documentation/stubs/qiskit.circuit.library.DraperQFTAdder.html}.
\newblock Accessed: 2022-01-03.

\bibitem{Bennett1984}
C.~H. Bennett and G.~Brassard, ``Quantum cryptography: Public key distribution
  and coin tossing,'' in {\em Proceedings of the IEEE International Conference
  on Computers, Systems, and Signal Processing}, pp.~175--179, {IEEE} Computer
  Society Press, 1984.

\bibitem{Bennett2014}
C.~H. Bennett and G.~Brassard, ``Quantum cryptography: Public key distribution
  and coin tossing,'' {\em Theoretical Computer Science}, vol.~560, pp.~7--11,
  2014.

\bibitem{Ampatzis2021}
M.~Ampatzis and T.~Andronikos, ``{QKD} based on symmetric entangled
  bernstein-vazirani,'' {\em Entropy}, vol.~23, no.~7, p.~870, 2021.

\bibitem{Aharon2010}
N.~Aharon and J.~Silman, ``Quantum dice rolling: a multi-outcome generalization
  of quantum coin flipping,'' {\em New Journal of Physics}, vol.~12, no.~3,
  p.~033027, 2010.

\bibitem{Meyer1999}
D.~A. Meyer, ``Quantum strategies,'' {\em Physical Review Letters}, vol.~82,
  no.~5, p.~1052, 1999.

\bibitem{Eisert1999}
J.~Eisert, M.~Wilkens, and M.~Lewenstein, ``Quantum games and quantum
  strategies,'' {\em Physical Review Letters}, vol.~83, no.~15, p.~3077, 1999.

\bibitem{Andronikos2018}
T.~Andronikos, A.~Sirokofskich, K.~Kastampolidou, M.~Varvouzou, K.~Giannakis,
  and A.~Singh, ``Finite automata capturing winning sequences for all possible
  variants of the {PQ} penny flip game,'' {\em Mathematics}, vol.~6, p.~20, Feb
  2018.

\bibitem{Andronikos2021}
T.~Andronikos and A.~Sirokofskich, ``The connection between the {PQ} penny flip
  game and the dihedral groups,'' {\em Mathematics}, vol.~9, no.~10, p.~1115,
  2021.

\bibitem{Giannakis2019}
K.~Giannakis, G.~Theocharopoulou, C.~Papalitsas, S.~Fanarioti, and
  T.~Andronikos, ``Quantum conditional strategies and automata for prisoners'
  dilemmata under the {EWL} scheme,'' {\em Applied Sciences}, vol.~9, p.~2635,
  Jun 2019.

\bibitem{Giannakis2015a}
K.~Giannakis, C.~Papalitsas, K.~Kastampolidou, A.~Singh, and T.~Andronikos,
  ``Dominant strategies of quantum games on quantum periodic automata,'' {\em
  Computation}, vol.~3, pp.~586--599, nov 2015.

\bibitem{Kastampolidou2020}
K.~Kastampolidou and T.~Andronikos, ``A survey of evolutionary games in
  biology,'' in {\em Advances in Experimental Medicine and Biology},
  pp.~253--261, Springer International Publishing, 2020.

\bibitem{Kastampolidou2021}
K.~Kastampolidou and T.~Andronikos, ``Microbes and the games they play,'' in
  {\em {GeNeDis} 2020}, pp.~265--271, Springer International Publishing, 2021.

\bibitem{Papalitsas2021}
C.~Papalitsas, K.~Kastampolidou, and T.~Andronikos, ``Nature and
  quantum-inspired procedures {\textendash} a short literature review,'' in
  {\em {GeNeDis} 2020}, pp.~129--133, Springer International Publishing, 2021.

\bibitem{Kastampolidou2022}
K.~Kastampolidou and T.~Andronikos, ``Game theory and other unconventional
  approaches to biological systems,'' in {\em Handbook of Computational
  Neurodegeneration}, pp.~1--18, Springer International Publishing, 2021.

\bibitem{Theocharopoulou2019}
G.~Theocharopoulou, K.~Giannakis, C.~Papalitsas, S.~Fanarioti, and
  T.~Andronikos, ``Elements of game theory in a bio-inspired model of
  computation,'' in {\em 2019 10th International Conference on Information,
  Intelligence, Systems and Applications ({IISA})}, {IEEE}, jul 2019.

\bibitem{Kastampolidou2020a}
K.~Kastampolidou, M.~N. Nikiforos, and T.~Andronikos, ``A brief survey of the
  prisoners' dilemma game and its potential use in biology,'' in {\em Advances
  in Experimental Medicine and Biology}, pp.~315--322, Springer International
  Publishing, 2020.

\bibitem{Kostadimas2021}
D.~Kostadimas, K.~Kastampolidou, and T.~Andronikos, ``Correlation of biological
  and computer viruses through evolutionary game theory,'' in {\em 2021 16th
  International Workshop on Semantic and Social Media Adaptation {\&}
  Personalization ({SMAP})}, {IEEE}, 2021.

\bibitem{Maschler2020}
M.~Maschler, {\em Game Theory}.
\newblock Cambridge University Press, 2020.

\bibitem{Dixit2015}
A.~Dixit, {\em Games of strategy}.
\newblock New York: W.W. Norton \& Company, 2015.

\bibitem{Tadelis2013}
S.~Tadelis, {\em Game Theory: An Introduction}.
\newblock Princeton University Press, 2013.

\end{thebibliography}

\end{document}